\documentclass[reprint,
superscriptaddress,
 amsmath,amssymb,
notitlepage,
prl,
10pt, 
tightenlines
]{revtex4-2}

\usepackage[utf8]{inputenc}
\usepackage[caption=false]{subfig}
\usepackage{graphicx}
\usepackage{dcolumn}
\usepackage{bm}
\usepackage{xcolor}
\usepackage{verbatim}
\usepackage{tabularx}
\usepackage{bbold}
\usepackage{float}
\usepackage{bm}
\usepackage{bbm}
\usepackage[colorlinks=True, citecolor=blue, urlcolor=blue]{hyperref}
\usepackage{booktabs}
\usepackage{natbib}

\usepackage{tikz}
\usetikzlibrary{quantikz}

\definecolor{crimson}{rgb}{.8, 0, 0}

\newtheorem{definition}{Definition}
\newcommand{\vp}{\varphi}

\begin{document}
\title{Quantum eigenstate broadcasting assisted by a coherent link}

\author{Benjamin~F.~Schiffer}
\email[Corresponding author: ]{Benjamin.Schiffer@mpq.mpg.de}
\affiliation{Max-Planck-Institut f\"ur Quantenoptik, Hans-Kopfermann-Str.~1, D-85748 Garching, Germany}%
\affiliation{Munich Center for Quantum Science and Technology (MCQST), Schellingstr.~4, D-80799 Munich, Germany}%
\author{Jordi~Tura}%
\affiliation{\small Instituut-Lorentz, Universiteit Leiden, P.O. Box 9506, 2300 RA Leiden, The Netherlands}%

\date{\today}

\begin{abstract}
Preparing the ground state of a local Hamiltonian is a crucial problem in understanding quantum many-body systems, with applications in a variety of physics fields and connections to combinatorial optimization. While various quantum algorithms exist which can prepare the ground state with high precision and provable guarantees from an initial approximation, current devices are limited to shallow circuits. Here we consider the setting where Alice and Bob, in a distributed quantum computing architecture, want to prepare the same Hamiltonian eigenstate. We demonstrate that the circuit depth of the eigenstate preparation algorithm can be reduced when the devices can share limited entanglement. Especially so in the case where one of them has a near-perfect eigenstate, which is more efficiently broadcast to the other device. Our approach requires only a single auxiliary qubit per device to be entangled with the outside. We show that, in the near-convergent regime, the average relative suppression of unwanted amplitudes is improved to $1/(2\sqrt{e}) \approx 0.30$ per run of the protocol, outperforming the average relative suppression of $1/e\approx 0.37$ achieved with a single device alone for the same protocol.
\end{abstract}

\maketitle

\emph{Introduction.}~--- 
Eigenstates, particularly ground states, play a crucial role in understanding physical models and solving problems in combinatorial optimization. As preparing the ground state of a local Hamiltonian is QMA-complete~\cite{kempe2006complexity} in general, no efficient quantum algorithm is expected to exist. However, preparing the ground state on a quantum computer still offers significant advantages, as the representation of a quantum many-body ground state already comes with an exponential advantage in memory over classical computers. 

Different quantum algorithms have been proposed so far for preparing ground states. This includes the quantum phase estimation (QPE) algorithm~\cite{Nielsen2016Quantum} as well as approximate algorithms such as the quantum adiabatic algorithm (QAA)~\cite{Messiah1962Quantum, Farhi2000Quantum}, filtering algorithms~\cite{Poulin2009Preparing, Ge2019Faster, Lu2021Algorithms} or variational algorithms~\cite{Cerezo2021Variational}. The QAA involves preparing a trivial ground state of an initial Hamiltonian and then sufficiently slowly changing the Hamiltonian into the target Hamiltonian. The runtime of the QAA depends polynomially on the inverse of the minimal spectral gap between the ground state and the first excited state along the adiabatic path and the final error~\cite{Jansen2007Bounds, Wiebe2012Improved}. 
Variational algorithmic methods have become increasingly popular in recent years. Performance guarantees for these methods are hard to obtain due to their heuristic nature and, in practice, the expressivity of the quantum circuit as well as (noisy) barren plateaus~\cite{Wang2020Noise} in the training process may hinder their performance.
The QPE remains a highly relevant algorithm for preparing the ground state precisely and variants of it have been investigated that are more suitable for early quantum devices where, e.g., only a single auxiliary qubit is required~\cite{OBrien2019Quantum, Santagati2018Witnessing}. These protocols focus on evaluating the eigenstate energy, which is closely connected to preparing the eigenstate itself.

\begin{figure}[b]
    \centering
    \includegraphics[width=1\columnwidth]{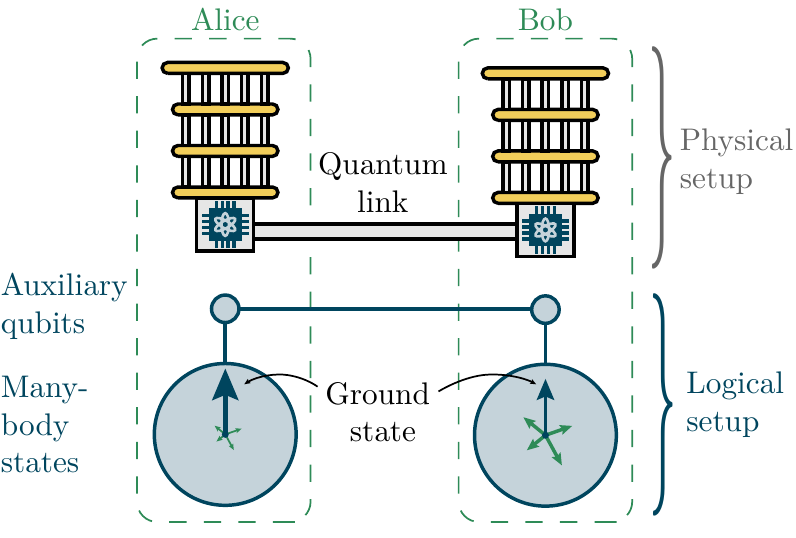}
    \caption{Illustration of the quantum broadcasting protocol with two quantum devices that are connected by a coherent quantum link which enables limited amount of entanglement sharing. In out setup, the device Alice is able to prepare a relatively good approximation (hence the larger ground state vector) to the ground state of $H$, while Bob can only prepare a more rudimentary approximation of the same state. By applying controlled quantum dynamics for random times $U=\exp(-iH\tau)$ and projecting into the fully symmetric subspace of the auxiliary qubits, which is heralded by the measurement outcome, the ground state on Bob can be prepared at arbitrary precision and much faster than without the help of Alice.}
    \label{fig:Fig1}
\end{figure}

Besides the limited coherence time of current devices, also the number of quantum bits (qubits) is modest. Indeed, increasing the qubit count appears to be a significant engineering challenge. An alternative pathway for more powerful quantum computers might lie in distributed quantum computing where separate quantum devices are able to share some entanglement. Recently, experiments with superconducting qubits have demonstrated a coherent microwave quantum link between spatially separated quantum devices~\cite{Magnard2020Microwave, harvey-collard_coherent_2022}. Motivated by such experimental progress, the question arises how the architecture may be exploited. A particularly relevant application of distributed quantum computing seems to be the problem of precise ground state preparation and verification. Without a quantum link, cross-platform verification could be performed where the desired state is prepared many times in order to perform randomized measurements where the worst-case complexity is exponential in the number of qubits~\cite{Elben2020Cross, Elben2022randomized}. Distributed quantum computing was investigated already in the context of communication between quantum devices~\cite{Cirac1999Distributed, Buhrman2001Quantum, buhrman_nonlocality_2010}.

A particularly interesting scenario that we shall consider here is where a quantum device, \emph{Alice}, is able to prepare a relatively good approximation to the ground state of a Hamiltonian, but another device, \emph{Bob}, is only able to prepare a much more rudimentary approximation of the same ground state (cf.~Fig.~\ref{fig:Fig1}).  Due to the no-cloning theorem~\cite{Nielsen2016Quantum}, it is not possible to simply copy the state from Alice to Bob. However, Alice can assist Bob in obtaining a more precise eigenstate.

In this letter we address this problem via a distributed quantum algorithm to prepare eigenstates of a Hamiltonian at arbitrary precision. Our protocol is inspired by the iterative quantum phase estimation algorithm and relies on a projection to the fully symmetric subspace of the auxiliary qubits that control the respective quantum dynamics.
Hereby, the devices share a limited amount of entanglement, in the sense that only the auxiliary qubits interact, which is suitable for experimental realizations~\cite{Monroe2014Large, Magnard2020Microwave, harvey-collard_coherent_2022}.
We show that the ability to share entanglement between the devices allows for a higher precision of the ground state prepared for Bob given a limited circuit.
In every iteration of the protocol, the successful subspace projection is heralded by the auxiliary qubits and approaches perfect success probability as the quantum state converges towards the desired eigenstate, resulting in a modest postselection overhead. 

In the limit of near-convergence to the ground state on Alice and Bob, we show that unwanted amplitudes for Bob are suppressed on average in every iteration by a factor of $1/(2\sqrt{e}) \approx 0.30$ relative to the ground state amplitude, in contrast to $1/e\approx 0.37$ when Alice is absent. Our analysis is complemented by numerical simulations demonstrating an advantage of the distributed protocol also in the regime before near-convergence to the ground state. We extend the discussion of the algorithm to the scenario where the ground state preparation is sped up by multiple coherently linked devices, each but one having access to a near-perfect ground state, where the subspace projection is related to a quantum Schur transformation on the auxiliary qubits~\cite{Bacon2006Efficient, Krovi2019efficient}. 

\begin{figure}[t]
    \centering
    \includegraphics[width=1\columnwidth]{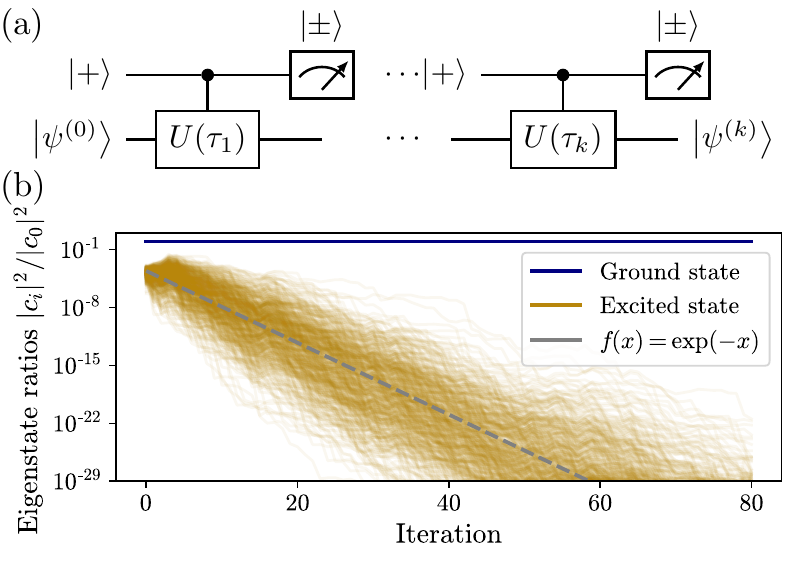}
    \caption{(a)~Quantum circuit for preparing a dominant eigenstate of $H$ with controlled dynamics $U(\tau_i)=\exp(-iH\tau_i)$ at every iteration.  The random $\{\tau_i\}$ are proportional to the inverse of the energy gap separating the the desired eigenstate in the spectrum. After the controlled dynamics, the auxiliary qubit is measured in the $\sigma^x$ basis. (b)~The single copy protocol is simulated for up to $k=100$ steps, on the ZZXZ model $H=\sum_{i=1}^{n=9} (\sigma_i^z \sigma_{i+1}^z + \sigma_i^x + \sigma_i^z)$. Here, a postselected instance is shown where the protocol successfully prepares the ground state. The protocol quickly converges to the ground state and the relative suppression of the subdominant amplitudes follows a trend with average $f(x)=\exp(-x)$.}
    \label{fig:Fig2}
\end{figure} 

\emph{Ground state preparation on a single device.}~--- 
First, we introduce our main toolset by analyzing iterative QPE in a single-device protocol. In this iterative QPE, we seek to prepare an eigenstate of a Hamiltonian $H$ at arbitrary precision. An initial approximation $\ket{\psi^{(0)}}$ to the eigenstate is prepared, e.g.~by adiabatic state preparation~\cite{Farhi2000Quantum}. Then, in every iteration of the protocol, the system is evolved with $H$ for a time $\tau_k$ while being controlled by an auxiliary qubit in the $\ket{+}$ state. We define $U_k = U(\tau_k)= \exp(-iH\tau_k)$. Finally, the auxiliary qubit is measured in the $\sigma^x$-basis. The measurement projects the state at the $k$-th iteration of the protocol from $\ket{\psi^{(k-1)}}$ to the (unnormalized) state $(I+U_{k})\ket{\psi^{(k-1)}}$ when a ``0" bit is obtained, otherwise to $(I-U_{k})\ket{\psi^{(k-1)}}$. Importantly, the evolution time $\tau_k$ is chosen at random for every iteration in a sufficiently large interval $\tau_k \in [0,\mathcal{O}(1/\Delta)]$ where $\Delta$ is the minimal energy gap separating the desired state from other eigenstates in the spectrum. The protocol is repeated and the former output state becomes the new input state [Fig.~\ref{fig:Fig2}(a)].

It is a well-known result that the canonical QPE may be used to project a given quantum state into an eigenstate of a Hamiltonian~\cite{Abrams_1999}, including iterative QPE~\cite{OBrien2019Quantum}. Here, we analyze the scaling behavior of eigenstate preparation of the iterative QPE, where the initial state is an approximation to the target state. We write the state before the $(k+1)$-th iteration of the protocol in the eigenbasis of $H$ as $\ket{\psi^{(k)}} = \sum_i c_i^{(k)} \ket{\phi_i}$. Using a Bayesian approach, we show that, on average, every run of the circuit suppresses the ratio of any non-dominant amplitude and the dominant amplitude $|c_j^{(k)}|^2 / |c_d^{(k)}|^2 $. When the is close to 1, the suppression is very small, and as the ratio approaches 0, the average suppression factor converges to $1/e$ [Fig.~\ref{fig:Fig2}(b)]. Thus, the suppression of unwanted amplitudes in terms of the ratios is exponential in the number of steps $k$. Due to the probabilistic nature of the protocol, convergence to the dominant eigenstate is not always guaranteed but the auxiliary qubit measurement heralds success/failure. In the case where the protocol fails, an initially subdominant eigenstate becomes the new dominant eigenstate which henceforth is amplified on average. The total success probability can be derived analogously to QPE and it is given by $\gamma^2$ where the initial overlap with the target eigenstate is $\gamma = |c_d^{(0)}|$. We write the action of the circuit in iteration $k$ as a quantum channel $\rho \rightarrow (\rho + U_k \rho U_k^\dag)/2 =: \mathcal{E}_k(\rho)$.
Hence, the probability to obtain a particular eigenstate $H\ket{\phi_i}=\lambda_i \ket{\phi_i}$ is given as
\begin{align}
    P_i = \bra{\phi_i} \mathcal{E}_k \left( \cdots \mathcal{E}_1\left( |\psi^{(0)}\rangle \langle\psi^{(0)}|\right)\cdots\right) \ket{\phi_i} = |c_i|^2.
    \label{eqn:channel}
\end{align}
Clearly, an eigenstate of $H$ is a fixed point of the channel sequence $\mathcal{E}_k \circ \cdots \circ \mathcal{E}_1$. Other states are projected towards an eigenstate which resembles an iterative Born rule~(cf.~\cite{Xu2014Demon, Chen2020Quantum}). 

Intuitively, the protocols succeeds in preparing the dominant eigenstate because the measurement of the auxiliary qubit at each iteration is correlated with the random variables $\vp_j^{(k)}:=\lambda_j\tau_k$ by the probability
$P_{\ket{0}}^{(k)} =\sum_l|c_l^{(k-1)}|^2\cos^2 (\vp_l^{(k)}/2)$
and $P_{\ket{1}}^{(k)} = 1 -  P_{\ket{0}}^{(k)}$, see the Supplementary Material for the derivation. If there is a strongly dominant amplitude, e.g.~of the ground state $c_0^{(k)} \lesssim 1$ and, if a randomly drawn value $(\vp_0^{(k)}\;\text{mod}\;2\pi)$ lies close to zero, then the likelihood of observing the $\ket{0}$ on the auxiliary qubit increases. This measurement will imprint a relatively large phase factor proportional to $\cos^2(\vp_0^{(k)}/2)$ onto the ground state amplitudes. However, the other non-dominant contributions $c_j^{(k)}$, $j\neq0$, are more likely to get a smaller phase factor because their respective $\vp_j^{(k)}$ correlates more weakly with the measured bit. The argument is formalized in the Supplementary Material and gives an average relative suppression of $1/e$ in the near-convergent regime. This method can, in principle, be used to prepare arbitrary eigenstates. For local Hamiltonians, some eigenenergies must become exponentially close. For these, our algorithm would behave like a filter at a random energy. That is the reason we focus on the ground state, where the energy gap is lower bounded by a constant in gapped phases and at critically it may vanish inverse polynomially with the system size. 

\begin{figure}[t]
    \centering
    \includegraphics[width=1\columnwidth]{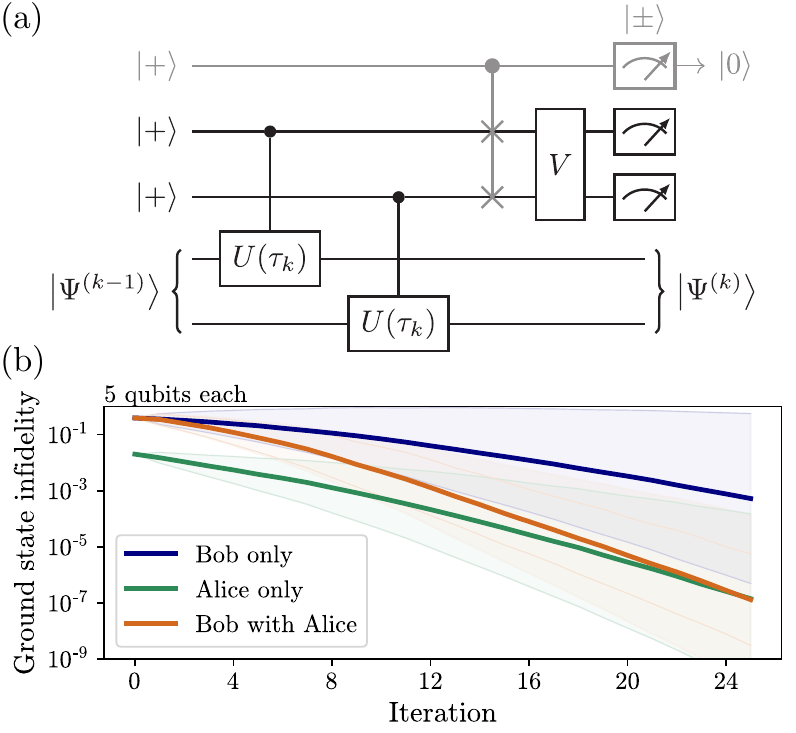}
    \caption{(a)~The quantum circuit for the protocol involving Alice with a good approximation to the ground state and Bob seeking to prepare the ground state at arbitrary precision from a more meager approximation to the ground state. The grayed part of the circuit (first wire and controlled swap) projects to the symmetric subspace and is only required for one of the two variants for quantum broadcasting. (b)~Numerical example of the quantum broadcasting protocol, postselected average over successful projections to the symmetric subspace. Shaded area shows one standard deviation of the data. The convergence speed for Bob is significantly higher when Alice is helping via the quantum link. We used the same Hamiltonian as in Fig.~\ref{fig:Fig2}(b).}
    \label{fig:Fig4}
\end{figure}

\emph{Quantum broadcasting with Alice and Bob.}~--- 
Our main result is a distributed quantum algorithm for broadcasting a quantum eigenstate from one device (Alice) to another (Bob). The protocol requires limited entanglement sharing: only one auxiliary qubit in each device needs to interact with only one qubit in the other. The setup is schematically depicted in Fig.~\ref{fig:Fig1} and the circuit shown in Fig.~\ref{fig:Fig4}(a). After identical controlled dynamics on both Alice and Bob, a joint operation on both auxiliary qubits entangles the states on the two devices. 
We consider two variants of the protocol as they offer different advantages. The first variant (depicted in black in [Fig.~\ref{fig:Fig4}(a)]) applies a Bell-basis transformation on the auxiliary qubits by choosing $V = [(1,0,0,1),(0,1,1,0),(0,1,-1,0),(1,0,0,-1)]/\sqrt{2}$.
In the second variant (including the gray part in [Fig.~\ref{fig:Fig4}(a)]) we make use of a controlled-SWAP operation and then apply Hadamard gates on the auxiliary qubits ($V=H\otimes H$) to measure in the $\sigma_x$-basis. A successful projection of the auxiliary qubits onto the symmetric space is heralded by the first helper qubit. In the limit of convergence to the ground state, the success probability $P_\text{sym}^{(k)}$ to project to the fully-symmetric subspace in the $k$-th iteration approaches $\lim_{k\rightarrow\infty} P_\text{sym}^{(k)} = 1$. The total postselection overhead is therefore modest if the initial overlap $\gamma$ is large and we estimate it to scale as $\mathcal{O}(1/\gamma^2)$ for variant 1 of the broadcasting setup. Given that current devices allow for rather shallow circuits only, this protocol allows to trade off circuit depth for heralded repetitions, for a target precision.

In Fig.~\ref{fig:Fig4}(b) we shows the average performance of the quantum broadcasting protocol for the variant $V=H\otimes H$, postselected to successful projection to the fully symmetric subspace. We compare the convergence speed for Alice and Bob separately and with the protocol presented. The convergence speed to the ground state for Bob is significantly higher due to the entanglement-enabled coordination with Alice. For simplicity, let us assume that Alice has a perfect ground state ($|c_{A,0}|=1$) while Bob's approximation is approaching $|c_{B,0}|\rightarrow 1$. Then, the average relative suppression of unwanted amplitudes $|c_{B,j}|$, $j>0$, postselected for successful projection to the symmetric subspace, is slightly weaker than for variant~1, amounting to $1/e^{9/8}\approx 0.32$ per round in the near-convergent limit. Beyond the near-convergent regime, our numerical studies show that, when Alice has a perfect eigenstate, variant~1 outperforms variant~2. In practice, though, Alice may only have a good approximation. In the latter case, numerics suggest that the performance of variant~1 decreases with the difference of fidelities in orders of magnitude, and eventually variant~2 becomes better when the fidelities become comparable, as we can observe in [Fig.~\ref{fig:Fig4}(b)].

Beyond the broadcasting scenario, the cross-platform verification allows for Alice and Bob to converge to the same eigenstate $\ket{\phi_i}\ket{\phi_i}$ in a heralded way. To this end, we just need two identical initial approximations $\ket{\psi}\ket{\psi}$. The same circuit as in Fig.~\ref{fig:Fig4}(a) can be used to this effect (see Supplementary Material). Naturally, if $\ket{\psi} \approx \ket{\phi_i}$ the protocol is more likely to produce two copies of the $i$-th eigenstate. This may be of particular interest in the absence of a suitable quantum memory.

An interesting adaptation is found by using $\tilde V = [(\sqrt{2},0,0,0),(0,1,1,0),(0,1,-1,0),(0,0,0,\sqrt{2})]/\sqrt{2}$, which produces a Schmidt decomposition in the eigenbasis of $U$:
$\ket{\Psi} = \sum_{i} c_{AB,ii} \ket{\phi_i}\ket{\phi_i}$, which might find interest in other quantum algorithms or e.g.~in the context of thermofield double states~\cite{Cottrell2019How}.

\emph{Extension to a multi-device setup.}~--- 
We consider also the extension to $p$ Alices with very good approximations to the same eigenstate and a single Bob. A natural generalization is to project all the auxiliary qubits to the symmetric space. 
For two devices, this projection can be implemented in a straightforward way (via Fig.~\ref{fig:Fig4}(a) or by performing a controlled swap operation on the two auxiliary qubits such that a measurement of the extra helper qubit signals projection into the symmetric or anti-symmetric subspace). For more devices, the projection to the fully symmetric subspace is related to the quantum Schur transform, which can be implemented efficiently~\cite{Bacon2006Efficient}. 

We seek to understand the scaling of our method in the limit of large $p$. The largest relative suppression seems to be present for the bit strings ``00\dots0" and ``11\dots1" which also allow for an analytical treatment. For this choice, which we conjecture to be optimal, we show that the relative suppression asymptotically approaches $1/4$. This coincides with the value for the bit string ``01" in the first variant for only two devices, suggesting a limit for how much the convergence to the ground state can be accelerated with the protocols discussed here. For the calculation, we refer to the Supplementary Material.

\begin{figure}[t]
    \centering
    \includegraphics[width=1\columnwidth]{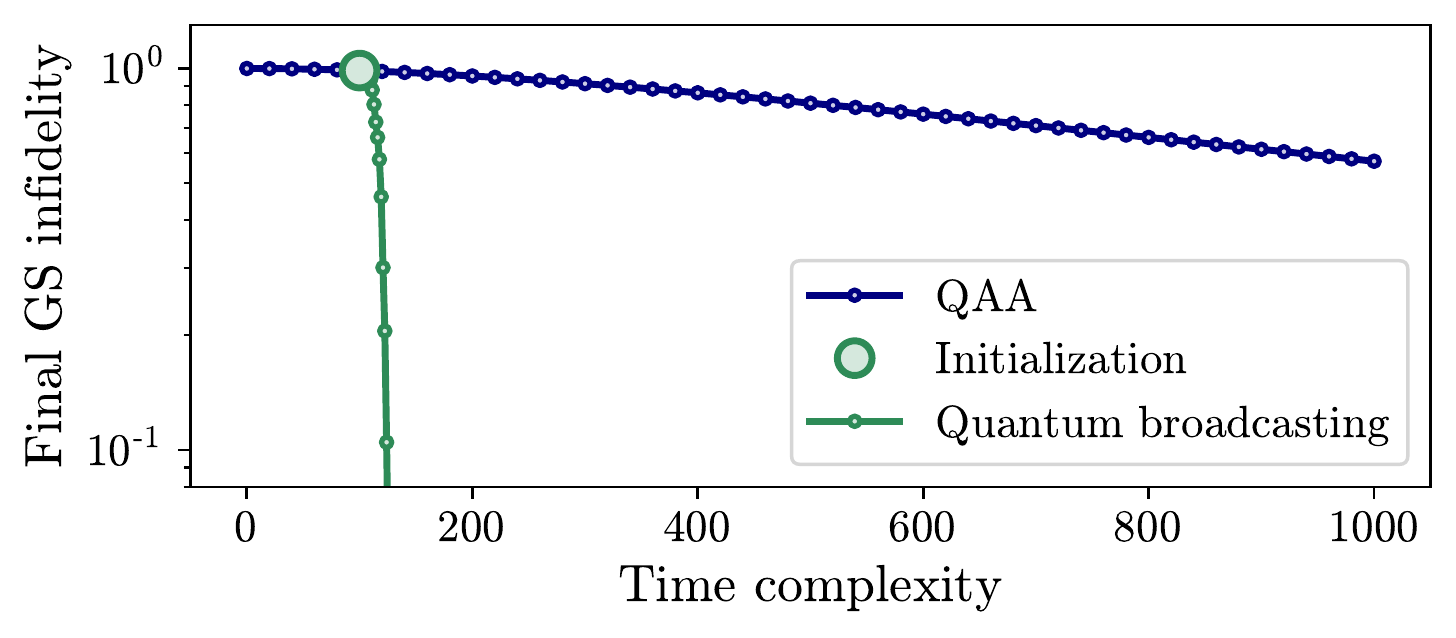}
    \caption{Combination of the quantum broadcasting protocol with the QAA for an instance of the ZZXZ model with 5 qubits. To prepare the ground state with the desired precision the adiabatic sweep may be performed slower. Alternatively, we suggest that an inferior approximation is used to initialize the QPE-inspired quantum broadcasting protocol. When the minimal adiabatic gap does not occur at the end of the sweep, the cooling algorithm significantly outperforms a QAA of longer evolution time, just considering the time complexity of the algorithm.}
    \label{fig:Fig5}
\end{figure}

\emph{Applications.}~--- 
The methods we here proposed prepare the ground state precisely from an approximation. This approximation may be obtained with the adiabatic algorithm~\cite{Farhi2000Quantum}. We may ask how the time in our methods compare with the additional time in the adiabatic algorithm for ground state preparation and show an example instance in Fig.~\ref{fig:Fig5}. Here, we used the ZZXZ model with a large interaction strength, such that the adiabatic gap is smaller than the spectral gap of the final Hamiltonian: $H_T=\sum_{i=1}^{n=9} (7\sigma_i^z \sigma_{i+1}^z + \sigma_i^x + \sigma_i^z)$. In the numerical simulation of the QAA we interpolate smoothly $H(s)=(s-1)H_0 + s H_T$ from a trivial Hamiltonian $H_0=\sum_{i=1}^{n=9} \sigma_i^z$ to observe an exponential error suppression in the regime considered. We note that the spectral gap for the target Hamiltonian for optimization problems cast into a ground state finding problem is often of constant size~\cite{Ebadi2022Quantum}.

Instead of using the cooling routines presented here at the end of an adiabatic sweep, one might also consider employing them at intermediate steps in the quasi-adiabatic evolution. Whenever the instantaneous state deviates too much from true ground state of the Hamiltonian (or possibly the ground state of a suitable corrected Hamiltonian~\cite{Benseny_2021}), detectable by non-destructive methods~\cite{Schiffer_2022}, several QPE iterations may significantly increase the instantaneous ground state overlap.

\emph{Discussion and Outlook.}~--- 
We have presented a distributed quantum protocol for ground state preparation at arbitrary precision. Compared to the iterative QPE, our protocol requires less repetitions to reach a certain ground state fidelity if a better approximation to ground state is available on a second quantum device and the devices are able share a limited amount of entanglement. We use a weaker assumption here than in the work by Koczor~\cite{Koczor2021Exponential} in that only the auxiliary qubits interact, which seems to be a more realistic scenario in line with experimental realizations~\cite{Magnard2020Microwave}.
The speedup for eigenstate preparation in the distributed protocol compared to the single device protocol is analytically derived for the near-convergent regime and beyond that we explore it qualitatively with numerics. A key advantage of the quantum broadcasting protocol is the built-in verification of the prepared state via heralding.

Our algorithm is inspired by QPE and relies on controlled quantum dynamics which are relatively costly to implement. Hence, the achieved reduction of depth in the distributed protocol is timely and relevant for early quantum devices. We note in this context that schemes have been proposed that replace controlled dynamics by a suitable clock or reference state~\cite{Lu2021Algorithms}. Also, neutral atom simulators enable conditional quantum dynamics without costly trotterized evolution~\cite{Bluvstein2021Controlling}.

In our protocol, the auxiliary qubits are measured, but the outcome is not used as we focus on preparing the ground state instead of learning the ground state energy. Future improvements may use the measured bits to gradually improve the choice for evolution time and measurement basis during the protocol.\\

The authors thank J.~Ignacio Cirac, Tom O'Brien and Dolev Bluvstein for insightful discussions. 
This work received funding from the European Union’s Horizon 2020 research and innovation program under Grant No.~899354 (FET Open SuperQuLAN) and through the ERC StG FINE-TEA-SQUAD (Grant No. 101040729).  This research is part of the Munich Quantum Valley, which is supported by the Bavarian state government with funds from the Hightech Agenda Bayern Plus.~J.T.~also acknowledges support from the Quantum Delta program. This publication is also part of the ‘‘Quantum Inspire – the Dutch Quantum Computer in the Cloud’’ project (with project number [NWA.1292.19.194]) of the NWA research program ‘‘Research on Routes by Consortia (ORC)’’, which is funded by the Netherlands Organization for Scientific Research (NWO). J.T. thanks the Google Research Scholar Program for support.

\let\oldaddcontentsline\addcontentsline
\renewcommand{\addcontentsline}[3]{}
\bibliography{biblio_paper}
\let\addcontentsline\oldaddcontentsline

\clearpage
\appendix
\onecolumngrid

\begin{center}
\textbf{\large Supplemental Materials: Quantum broadcasting of eigenstates by entanglement sharing}
\end{center}

\setcounter{secnumdepth}{2}
\setcounter{equation}{0}
\setcounter{figure}{0}
\setcounter{table}{0}
\setcounter{page}{1}
\makeatletter
\renewcommand{\theequation}{S\arabic{equation}}
\renewcommand{\thefigure}{S\arabic{figure}}
\renewcommand{\bibnumfmt}[1]{[S#1]}

\tableofcontents

\clearpage
\section{Eigenstate preparation on a single device} \label{sec:Alice}

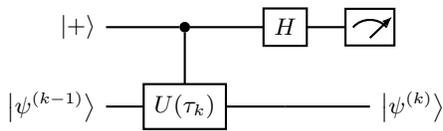
\begin{figure}[htbp]
    \centering
    \begin{tikzpicture}
        \node[scale=1] {
            \begin{quantikz}
                \lstick{\ket{+}}  &  \ctrl{1} &  \gate{H}  & \meter{} & \\
                \lstick{$\big|\psi^{(k-1)}\big\rangle$}  &  \gate{U(\tau_k)} &   \qw &  \qw\rstick{$\big|\psi^{(k)}\big\rangle$} & 
            \end{quantikz}
        };
    \end{tikzpicture}
    \caption{A quantum circuit, which is essentially the Hadamard test circuit, is applied to an initial state $\big|\psi^{(k-1)}\big\rangle$. The auxiliary qubit is then reset and the output state  $\big|\psi^{(k)}\big\rangle$ becomes the next initial state.}
    \label{fig:1copycircuit}
\end{figure}

We are interested in a class of QPE-inspired algorithms which prepare a particular eigenstate with a desired precision, given an initial approximation to such eigenstate. Here, we consider the single device case, depicted in Fig.~\ref{fig:1copycircuit}. 

\subsection{Quantum circuit and process map}

We begin by explaining how repeated application of the Hadamard circuit amplifies a dominant eigenstate on average.
The protocol requires the ability to implement controlled unitary evolution: an auxiliary qubit is initialized in the \ket{+} state, then the controlled-Hamiltonian evolution $U$ of $H$ is applied for a time $\tau_k$, which we denote simply by $U(\tau_k)$, and the auxiliary qubit is measured in the $\sigma_x$-basis. To analyze the behavior of a single run of the circuit, we write $U$ and the initial state in $\ket{\psi}$ in the eigenbasis of $H$:
\begin{align}
U(\tau_k)=e^{-i\tau_k H}=\sum_j e^{-i\tau_k\lambda_j} \ket{\phi_j}\bra{\phi_j} =  \sum_j e^{i \varphi_j} \ket{\phi_j}\bra{\phi_j},
\end{align}
where we have dropped the $k$ label to simplify the notation, and $\ket{\psi} = \sum_j c_j \ket{\phi_j}$ with $c_j\in \mathbb{C}$. The number of qubits of the initial state is $n$ (so the number of eigenstates is $N=2^n$). The composite system before the measurement can be described by
\begin{align}
    &\frac{1}{2}\left(\ket{0} \big( \ket{\psi} + U \ket{\psi} \big) + \ket{1}  \big( \ket{\psi} - U \ket{\psi} \big)  \right) \\
    =& \frac{1}{2}\left(\ket{0} \sum_j c_j (1+e^{i\vp_j})  \ket{\phi_j}  + \ket{1}  \sum_j c_j (1-e^{i\vp_j})  \ket{\phi_j}  \right).
\end{align}
We are interested in studying the following process: $|c_j^{(0)}|^2\mapsto |c_j^{(k)}|^2$, where $k$ denotes the current iteration. For our analysis we rely on the following assumptions:
\begin{itemize}
    \item[(1)] The circuit is noiseless.
    \item[(2)] The variables $\vp_j^{(k)}:=\lambda_j\tau_k \mod 2\pi$ are independent and identically distributed as $\vp_j^{(k)}\sim \mathrm{Unif}[0,2\pi]$ for all $j, k$.
\end{itemize}

To justify (2) we need to take, for each iteration, the evolution time $\tau_k$ as a uniform non-negative random variable such that $\tau_k\in O(1/\Delta)$. This guarantees that all the eigenphases are effectively randomized with respect to the ground state. Moreover, any statistical dependence between their respective eigenphases is removed so $\varphi_0$ and $\varphi_j$ are statistically independent. This follows from $\Delta$ being the spectral gap of $H$. Naturally, a very similar argument can be done for eigenstates above the ground state as long as one has a guarantee on the minimal distance between the eigenenergy we are interested in and the closest eigenenergy to it, which we can then take as $\Delta$. Furthermore, when we are interested in the ground state, the analysis will show that statistical dependence among higher values of $j$ becomes irrelevant since $|c_j|\ll 1$.
    We therefore have a family of random variables $\vp_j^{(k)}:=\lambda_j\tau_k \mod 2\pi \sim \mathrm{Unif}[0,2\pi]$ i.i.d.

We can define a process map that converts the vector of eigenstate populations $\{|c_j^{(k-1)}|^2\}_j$ onto the next vector of eigenstate populations $\{|c_j^{(k)}|^2\}_j$.
The process map for the $k$-th iteration can be described as
\begin{equation}\label{eq:Aprocessmap}
|c_j^{(k-1)}|^2 \mapsto |c_j^{(k)}|^2 = |c_j^{(k-1)}|^2\times \frac{S_(i,j)^{(k)}}{N^{(k)}} = |c_j^{(k-1)}|^2\times  \left\{
\begin{array}{ccc}
    \displaystyle\frac{(1+\cos(\vp_j^{(k)}))}{\displaystyle\sum_{l}|c_l^{(k-1)}|^2(1+\cos(\vp_l^{(k)}))} &  \mbox{with prob.} & \Pr{}_{\ket{0}}^{(k)},\\
    \displaystyle\frac{(1-\cos(\vp_j^{(k)}))}{\displaystyle\sum_{l}|c_l^{(k-1)}|^2(1-\cos(\vp_l^{(k)}))} &  \mbox{with prob.} & \Pr{}_{\ket{1}}^{(k)}.
\end{array}
\right.
\end{equation}
The $S_j^{(\alpha)}$ are the respective phase factors that are multiplied onto the population for the $j$-th eigenstate at iteration $(k-1)$ and $N^{(\alpha)}$ is the normalization. The probabilities to measure a $\ket{0}$ or $\ket{1}$ in the ancillary qubit are given as
\begin{equation} \label{eq:A_prob}
    \Pr{}_{\ket{0}}^{(\alpha)} =\sum_l\frac{|c_l^{(\alpha-1)}|^2}{2}(1 + \cos(\vp_l^{(\alpha)})), \quad \text{and} \quad 
    \Pr{}_{\ket{1}}^{(\alpha)} =\sum_l\frac{|c_l^{(\alpha-1)}|^2}{2}(1 - \cos(\vp_l^{(\alpha)})).
\end{equation}

The algorithm converges to an eigenstate (cf.~App.~\ref{ssec:symmetry}) of $H$. Here we are interested in quantifying the convergence speed to an eigenstate. For simplicity, we take in our analysis this eigenstate to be the ground state. This motivates the following definition:
\begin{definition}[Dominant eigenvector]\label{def:1}
For a Hamiltonian $H=\sum_i \lambda_i \ket{\phi_i}\bra{\phi_i}$ with non-degenerate spectrum and a quantum state $\ket{\psi}=\sum_i c_i \ket{\phi_i}$, $\ket{\psi}$ has a single dominant eigenvector $\ket{\phi_d}$ if and only if $|c_d|^2 > |c_i|^2$, $\forall i\neq d$. We define the dominance $D:= |c_d|^2 - |c_s|^2$ as a distance from the dominant eigenstate $\ket{\phi_d}$ to the subdominant eigenstate $\ket{\phi_s}$ where $|c_d|^2 > |c_s|^2 > |c_i|^2$, $\forall i \notin \{d,s\}$.
\end{definition}
In the algorithm, the eigenstate that we want to prepare is the dominant eigenstate with an initial amplitude $|c_d^{(0)}|^2$ and initial dominance $D^{(0)}$. 
After the first $k$ rounds of the protocol, the $|c_j^{(k)}|^2$ are random variables described by
\begin{align}
    |c_j^{(0)}|^2 \mapsto |c_j^{(k)}|^2 = |c_j^{(0)}|^2 \times \frac{S_j^{(1)} S_j^{(2)} \cdots S_j^{(k)}}{N^{(1)} N^{(2)} \cdots N^{(k)}}.
\end{align}

\subsection{Bayesian approach for the expected amplitude factor}

The $S_j^{(\alpha)}$ and the $N^{(\alpha)}$ can be considered as random variables. The $S_j^{(\alpha)}$ are independent of each other for every $\alpha, j$, according to assumption~(2). We note that the $|c_j^{(\alpha)}|^2$ depend only on the outcome of the previous round $\alpha-1$. Observe that we can describe the process as a discrete-time Markov chain in an infinite state space. 
For our analysis, we consider the ratios of two amplitudes at the same iteration so that we do not have to take into account the normalization
\begin{align} \label{eq:S_frac}
    \frac{|c_j^{(k)}|^2 }{|c_d^{(k)}|^2} = \frac{|c_j^{(0)}|^2 }{|c_d^{(0)}|^2} \times \frac{S_j^{(1)} S_j^{(2)} \cdots S_j^{(k)}}{S_d^{(1)} S_d^{(2)} \cdots S_d^{(k)}}. 
\end{align}
We use the index $d$ (of the dominant state) in the denominator, as we are interested in quantifying how quickly non-dominant amplitudes are suppressed. Taking the natural logarithm, we obtain
\begin{align}
    \log \left(\frac{|c_j^{(k)}|^2 }{|c_d^{(k)}|^2}\right) =   \log \left( \frac{|c_j^{(0)}|^2 }{|c_d^{(0)}|^2} \right) + \log \big(S_j^{(1)}\big) - \log \big(S_d^{(1)}\big) + \dots + \log \big(S_j^{(k)}\big) - \log \big(S_d^{(k)}\big). \label{eq:log(S)}
\end{align}

of near convergence to the ground state so that we can consider the expectation value of $\log \left(|c_j^{(k)}|^2 /|c_d^{(k)}|^2 \right)$ in this limit. We assume that $|c_d^{(\alpha)}|^2 \rightarrow 1$ and  $|c_j^{(\alpha)}|^2 \rightarrow 0$ for $j\neq d$, such that we consider the $S_j^{(\alpha)}$ (and $S_d^{(\alpha)}$) to both being independent and identically distributed. Then, from the central limit theorem we conclude that the $S_j^{(\alpha)}$ are log-normal distributed if the variance of $S_j^{(\alpha)}$ is finite. Hence, we seek to calculate the moments of $S_j^{(\alpha)}$.
We note that the $\vp_j^{(\alpha)}$ are $\vp_j^{(\alpha)} \sim \mathrm{Unif}[0,2\pi]$ i.i.d.~for every $\alpha$. Nevertheless, as the probability to obtain a ``0" or a ``1" depends on the $\vp_j^{(\alpha)}$, the probability distribution we have access to is actually the distribution of the $\vp_j^{(\alpha)}$ conditioned on the results of the auxiliary qubit. Hence, given that we observe, say, a ``0", the $\vp_j^{(\alpha)}$ are no longer uniform but become biased as a function of the coefficients $\{c_j\}$. 

Hence, we take a Bayesian approach to computing the moments $S_j^{(\alpha)}$ and begin by recalling Bayes' theorem in its most basic form:
\begin{align}
    p(A|B) = \frac{p(B|A)p(A)}{p(B)}.
\end{align}
Given that we observe a $\ket{0}$ on the auxiliary qubit and that the coefficients are $\{c_j^{(\alpha-1)}\}$, the probability density for a set $\vp_j^{(\alpha)}$ is given by
\begin{align} \label{eq:biasedvarphiAlice}
    p\Big(\vp_j^{(\alpha)} \Big| \ket{0}, \{c_j^{(\alpha-1)}\}\Big) 
    =& \frac{ p\Big(\ket{0} \Big| \vp_j^{(\alpha)} , \{c_j^{(\alpha-1)}\}\Big) \times p\Big(\vp_j^{(\alpha)} \Big| \{c_j^{(\alpha-1)}\}\Big)}{p\Big(\ket{0} \Big| \{c_j^{(\alpha-1)}\}\Big)}
\end{align}
which is simply Bayes' theorem applied to probability densities conditioned on $\{c_j^{(\alpha-1)}\}$. Observe that the probability density for a set $\vp_j^{(\alpha)}$ without conditioning on the measured auxiliary qubit is independent of the $\{c_j^{(\alpha-1)}\}$, because $\tau_k$ is chosen independently of the $\{c_j^{(\alpha-1)}\}$, so we have
\begin{align}
    p\Big(\vp_j^{(\alpha)} \Big| \{c_j^{(\alpha-1)}\}\Big) = p\Big(\vp_j^{(\alpha)} \Big) = \frac{1}{(2\pi)^N}.
\end{align}
In addition, the probability to measure a $\ket{0}$ given a set of $\{c_j^{(\alpha-1)}\}$ but not conditioning on the $\vp_j^{(\alpha)}$ is
\begin{align}
    p\Big(\ket{0} \Big| \{c_j^{(\alpha-1)}\}\Big) = \frac{1}{(2\pi)^N} \int_0^{2\pi} \cdots \int_0^{2\pi} p\Big(\ket{0} \Big| \vp_j^{(\alpha)} , \{c_j^{(\alpha-1)}\}\Big) d\vp_1^{(\alpha)}\dots d\vp_{N}^{(\alpha)} = \frac{1}{2}
\end{align}
where we used the probability previously stated
\begin{align}
    p\Big(\ket{0} \Big| \vp_j^{(\alpha)} , \{c_j^{(\alpha-1)}\}\Big) =  \Pr{}_{\ket{0}}^{(\alpha)}
\end{align}
in Eq.~\ref{eq:A_prob}. We can insert all three terms into Eq.~\ref{eq:biasedvarphiAlice} to obtain
\begin{align}
    p\Big(\vp_j^{(\alpha)} \Big| \ket{0}, \{c_j^{(\alpha-1)}\}\Big) 
    = \frac{1}{(2\pi)^N} \sum_l|c_l^{(\alpha-1)}|^2(1 + \cos(\vp_l^{(\alpha)})).
\end{align}

In order to apply the Central Limit Theorem to Eq.~\ref{eq:log(S)}, we compute the first and second moment of $\log \big[ S_j^{(\alpha)} \big]  = \log \big[1 + \cos(\vp_j^{(\alpha)})\big]$. The expectation value (over the $\vp$ variables) is given as
\begin{align}
    \mu_{j,\alpha} =& \mathbb{E}\left[\log \big[ S_j^{(\alpha)} \big] \Big| \ket{0}, \{c_j^{(\alpha-1)}\}\right] \\
    =& \frac{1}{(2\pi)^N} \int_0^{2\pi} \cdots \int_0^{2\pi} \log \big[1 + \cos(\vp_j^{(\alpha)})\big]  \left( \sum_l|c_l^{(\alpha-1)}|^2 \big[1 + \cos(\vp_j^{(\alpha)})\big] \right) d\vp_1^{(\alpha)}\dots d\vp_{N}^{(\alpha)} \\
    =& \frac{1}{(2\pi)} \int_0^{2\pi} \log \big[1 + \cos(\vp_j^{(\alpha)})\big]  \big[1+|c_j^{(\alpha-1)}|^2 \cos(\vp_j^{(\alpha)})\big] d\vp_j^{(\alpha)} \\
    =& |c_j^{(\alpha-1)}|^2 - \log(2) \label{eq:alicemoment}
\end{align}
and for the variance we need to compute
\begin{align}
    \sigma_{j,\alpha}^2 = \text{Var}\left[\log \big[ S_j^{(\alpha)} \big] \Big| \ket{0}, \{c_j^{(\alpha-1)}\}\right]
    =  \mathbb{E}\left[\log \big[ S_j^{(\alpha)} \big]^2 \Big| \ket{0}, \{c_j^{(\alpha-1)}\}\right] -  \mathbb{E}\left[\log \big[ S_j^{(\alpha)} \big] \Big| \ket{0}, \{c_j^{(\alpha-1)}\}\right] ^2.
\end{align}
We calculate
\begin{align}
    \mathbb{E}\left[\log \big[ S_j^{(\alpha)} \big]^2 \Big| \ket{0}, \{c_j^{(\alpha-1)}\}\right]
    =& \frac{1}{(2\pi)} \int_0^{2\pi}\log \big[1 + \cos(\vp_j^{(\alpha)})\big]^2   (1+|c_j^{(\alpha-1)}|^2 \cos(\vp_j^{(\alpha)})) d\vp_j^{(\alpha)} \\
    =& \frac{\pi^2}{3} + \log(2)^2 - |c_j^{(\alpha-1)}|^2 (2+\log(4))
\end{align}
and obtain for the variance
\begin{align}
   \sigma_{j,\alpha}^2 
    =  \frac{\pi^2}{3} - |c_j^{(\alpha-1)}|^2 \big(2+|c_j^{(\alpha-1)}|^2 \big) < \infty,
\end{align}
which is finite. 
Note that for the complementary case, when a $\ket{1}$ is measured, we obtain the same results for the relevant moments, due to symmetries of the trigonometric functions. Therefore, we write $\mu_{j,\alpha}$ and $ \sigma_{j,\alpha}^2 $ without an explicit dependency on the auxiliary qubit measurement result.

In order to apply the Central Limit Theorem, we require the $\log \big[ S_j^{(\alpha)}\big]$ to be independent and identically distributed. In the limit of near convergence to the ground state, both conditions hold. We denote by $k_0$ the iteration after which we enter in the near-convergence regime. For simplicity, we therefore consider the relevant moments of $\log \left(|c_j^{(\alpha)}|^2 /|c_d^{(\alpha)}|^2 \right)$ in this limit, which implies that we assume that $|c_d^{(\alpha)}|^2 \rightarrow 1$ and  $|c_j^{(\alpha)}|^2 \rightarrow 0$ for $j\neq d$ are approximately constant, i.e.~they also become independent of the previous iteration. In this case, the $S_j^{(\alpha)}$ are log-normal distributed if the variance of $S_j^{(\alpha)}$ is finite, which is the case.
We remark that this statement admits a broader range of applicability, since there exist extended versions of the Central Limit Theorem (such as Lyapunov's or Lyndeberg's) that hold under slightly more relaxed assumptions.

For a sequence of $m$ samples in the near-convergent limit $\{\dots,\log \big[ S_j^{(k_0)}\big], \dots, \log \big[ S_j^{(k_0+m)}\big]\}$, the sum of the samples by the central limit theorem (CLT) approaches a normal distribution 

\begin{align}
    \mathcal{LS}_j = \sum_{\alpha=k_0}^{k_0+m} \log \big[ S_j^{(\alpha)}\big] 
    \sim  \mathcal{N} \left(-m \log(2), m \pi^2/3\right)
\end{align}
for large $m$.
For the dominant amplitude, respectively, we have \begin{align}
    \mathcal{LS}_d = \sum_{\alpha=k_0}^{k_0+m} \log \big[ S_d^{(\alpha)}\big] 
    \sim  \mathcal{N} \left(-m (1-\log(2)), m (\pi^2/3-3)\right).
\end{align}
For reference, $-\log(2) \approx -0.693$ and $1-\log(2)\approx 0.307$; $\pi^2/3\approx 3.29$ and $\pi^2/3-3\approx 0.29$.

\subsection{Computing the average relative suppression}

In this section we are interested in the distribution of the ratio $|c_j^{(k_0+m)}|^2 /|c_d^{(k_0+m)}|^2$ with $j\neq d$, thus with Eq.~\ref{eq:S_frac} we first compute
\begin{align}
    \exp(\mathcal{LS}_j - \mathcal{LS}_d) \sim& \ \text{Lognormal}\left(-m, 
m (2\pi^2/3-3) \right).
\end{align}
We are interested in the average relative suppression per round in the near-convergent regime. The average suppression of subdominant amplitudes is the mean of the distribution
\begin{align}
\exp(\mathcal{LS}_j - \mathcal{LS}_d)^{1/m} \sim& \ \text{Lognormal}\left(-1, (2\pi^2/3-3)/m\right).
\end{align}

As the mean of a lognormal distribution $\text{Lognormal}(\mu, \sigma^2)$ is given by $\exp(\mu + \sigma^2/2)$ we can now write the expected suppression as
\begin{align}
      \lim_{m\rightarrow\infty} \exp \left( -1 +  \frac{2\pi^2/3-3}{2m} \right) = \exp(-1). \label{eq:variancevanishes}
\end{align}

Thus, we derived the expected suppression of a non-dominant eigenstate $\ket{\phi_j}$ relative to the dominant eigenstate $\ket{\phi_d}$ as a function of the difference of their respective amplitudes. Naturally, the expected suppression is the weakest for the pair of the subdominant and dominant eigenstates.
Qualitatively, we expect the ratio to be suppressed in a stronger way, the more the subdominant amplitude is separated from the dominant amplitude (cf.~Eq.~\ref{eq:alicemoment}). 
We note that there exists the possibility of two eigenstates crossing, i.e.~the initially dominant amplitude becoming the new subdominant eigenstate, which reduces the likelihood of preparing the initially dominant eigenstate. This is discussed in the main text and it is a consequence the Born rule.

\clearpage
\section{Quantum broadcasting: 2-auxiliary qubit variant} \label{sec:twoancillas}

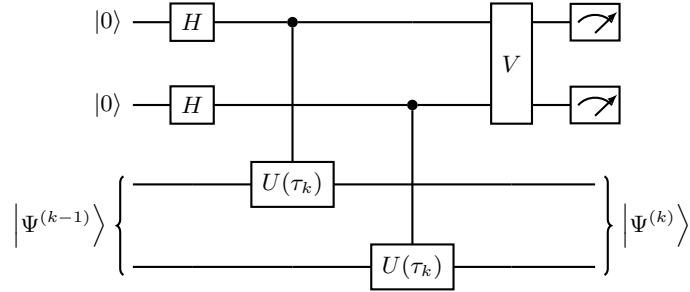
\begin{figure}[h]
    \centering
    \begin{tikzpicture}
    \node[scale=1] {
        \begin{quantikz}
             \lstick{\ket{0}} &  \gate{H}  &  \ctrl{2} &  \qw &  \gate[wires=2]{V} &  \meter{} & \\
             \lstick{\ket{0}} &  \gate{H}  & \qw  &  \ctrl{2}   &  & \meter{}& \\
            \lstick[wires=2]{$\ket{\Psi^{(k-1)}}$}  & \qw &  \gate{U(\tau_k)} &   \qw &  \qw &   \qw\rstick[wires=2]{$\ket{\Psi^{(k)}}$} & \\
              &  \qw &  \qw &   \gate{U(\tau_k)} &    \qw &  \qw  &
        \end{quantikz}
        };
    \end{tikzpicture}%
    \caption{Quantum broadcasting circuit with two auxiliary qubits controlling the respective dynamics on two quantum devices. The auxiliary qubits are initialized in the $\ket{0}$ state followed by a local Hadamard transformation into $\ket{+}$ respectively. The measurements are performed in the computational basis.}
    \label{fig:maincircuit}
\end{figure}

Here we calculate the relative average suppression for the quantum broadcasting protocol with two auxiliary qubits in the same spirit as for the single device algorithm. The circuit for one iteration of the protocol is shown in Fig.~\ref{fig:maincircuit} with the operator $V$ defined as
\begin{align}
    V = \frac{1}{\sqrt{2}} \begin{pmatrix}
    1 & 0 & 0 & 1 \\
    0 & 1 & 1 & 0 \\
    0 & 1 & -1 & 0 \\
    1 & 0 & 0 & -1 
    \end{pmatrix}.
\end{align}

\subsection{Quantum circuit and process map}

First, we write down the action of the quantum circuit for a general $V$. We define $\ket{\psi'}:=U\ket{\psi}$. To build an intuition, we begin considering the case where $\ket{\Psi^{(0)}} = \ket{\psi}\ket{\psi}$ and consider a single iteration of the circuit in Fig. S2. Before the computational basis measurements, the global state of the circuit is
\begin{align}  \label{eq:withV}
 \frac{1}{2} \Big[ V\ket{00} \ket{\psi \psi} + V\ket{01} \ket{\psi \psi'} +  V\ket{10} \ket{\psi' \psi} + V\ket{11} \ket{\psi' \psi'} \Big].
\end{align}
Substituting the operator $V$ by our definition, we obtain
\begin{align}
    & \frac{1}{2\sqrt{2}} \Big[ \ket{00} \big( \ket{\psi \psi} + \ket{\psi' \psi'} \big)  + \ket{01}  \big( \ket{\psi \psi'} + \ket{\psi' \psi} \big)  +  \ket{11} \big( \ket{\psi \psi} - \ket{\psi' \psi'}\big) + \ket{10}  \big( \ket{\psi' \psi} - \ket{\psi \psi'} \big)  \Big] \\
    =& \frac{1}{2\sqrt{2}} \Big[ \ket{00} \sum_{i,j} c_ic_j (1+ e^{i(\vp_i + \vp_j)}) \ket{\phi_i \phi_j} \nonumber
    + \ket{01} \sum_{i,j} c_ic_j (e^{i\vp_i} + e^{i\vp_j})\ket{\phi_i \phi_j}  \\ 
    &+ \ket{11} \sum_{i,j} c_ic_j (1 - e^{i(\vp_i + \vp_j)}) \ket{\phi_i \phi_j} ) + \ket{10} \sum_{i,j} c_ic_j (e^{i\vp_i} - e^{i\vp_j})\ket{\phi_i \phi_j}  \Big]. \label{eq:circuitAB}
\end{align}
In the general case where the input is $\ket{\Psi^{(k-1)}}$, the states held by Alice and Bob may be entangled in general, which means that we have coefficients $c_{ij}$ instead of $c_i c_j$ in Eq.~\ref{eq:circuitAB}.
We already see that the probability to measure $\ket{10}$ on the auxiliary qubits approaches zero as the algorithm converges towards an identical eigenstate pair $\ket{\Psi} \approx \ket{\phi_j}\ket{\phi_j}$. We analyze the postselection overhead in Sec.~\ref{ssec:post}. When a certain bit string is observed on the auxiliary qubits, the 
amplitudes are multiplied by their respective local phase factors. The corresponding process map for the $k$-th iteration is given by
\begin{align}
    |c_{ij}^{(k-1)}|^2\mapsto  |c_{ij}^{(k)}|^2 =&  |c_{ij}^{(k-1)}|^2\times \frac{S_{(i,j),\ket{b}}^{(k)}}{N^{(k)}_{\ket{b}}} \\
    =& |c_{ij}^{(k-1)}|^2\times \left\{
    \begin{array}{ccc}
        \big[1+\cos(\vp_i^{(k)}+\vp_j^{(k)})\big]/\big( 4 N_{\ket{00}}^{(k)} \big) &  \mbox{with} & \Pr{}_{\ket{00}}^{(k)},\\[10pt]
        \big[1+\cos(\vp_i^{(k)}-\vp_j^{(k)})\big]/\big( 4 N_{\ket{01}}^{(k)} \big) &  \mbox{with} & \Pr{}_{\ket{01}}^{(k)},\\[10pt]
        \big[1-\cos(\vp_i^{(k)}+\vp_j^{(k)})\big]/\big( 4 N_{\ket{11}}^{(k)} \big) &  \mbox{with} & \Pr{}_{\ket{11}}^{(k)},\\[10pt]
        \big[1-\cos(\vp_i^{(k)}-\vp_j^{(k)})\big]/\big( 4 N_{\ket{10}}^{(k)} \big) &  \mbox{with} & \Pr{}_{\ket{10}}^{(k)}, \label{eq:ABprocessmap}
    \end{array}
    \right.    
\end{align}
where we have made more explicit the dependency of $S$ with the auxiliary qubit measurement result (cf.~Eq.~\ref{eq:Aprocessmap}), which is denoted by the subscript $\ket{b}$.
For convenience, we absorb the factor of $1/4$ in the $S$ into the normalization $N$ so we can redefine $S_{(i,j),\ket{b}}^{(k)} := 1 \pm \cos(\vp_i^{(k)} \pm \vp_j^{(k)})$ with the signs chosen as in Eq.~\ref{eq:ABprocessmap}.
Note that the probabilities $\Pr{}_{\ket{b}}^{(\alpha)} = [1 \pm \cos(\vp_i^{(k)} \pm \vp_j^{(k)})]/4$ add up to one: $\sum_{b \in \{0,1\}^2} \Pr{}_{\ket{b}}^{(\alpha)}$ = 1.

\subsection{Bayesian approach for the expected amplitude factor}

As in the single device case, we consider the $S_{(i,j),\ket{b}}^{(\alpha)}$ as random variables, where the measurement outcome bit string is denoted as $b$.
We write the probability to measure a certain bit string $b$ now as $\Pr{}_{\ket{b}}^{(\alpha)} = p\Big(\ket{b} \Big| \vp_i^{(\alpha)}, \{c_{ij}^{(\alpha-1)}\}\Big)$, i.e.~the probability depends on the coefficients $\{c_{ij}^{(\alpha-1)}\}$ and the product of the evolution time with the eigenenergies $\vp_i^{(\alpha)}$ (note that these are the same for Alice and Bob since both are evolving for the same $\tau_k$). We list, in the same spirit as for the single device case, the probabilities of observing a bit string given a set of coefficients
\begin{align}
    & p\Big(\ket{00} \Big| \{c_{ij}^{(\alpha-1)}\}\Big) = p\Big(\ket{11} \Big| \{c_{ij}^{(\alpha-1)}\}\Big) =  \frac{1}{4},\\
    & p\Big(\ket{01} \Big| \{c_{ij}^{(\alpha-1)}\}\Big) = \frac{1}{4}  + \frac{1}{4} \sum_i |c_{ii}^{(\alpha-1)}|^2, \\
    & p\Big(\ket{10} \Big| \{c_{ij}^{(\alpha-1)}\}\Big) = \frac{1}{4} \sum_{i\neq j} |c_{ij}^{(\alpha-1)}|^2. \label{eq:AB_p10}
\end{align}
Also, we note the a priori probability density to observe a $\vp_i^{(\alpha)}$ is
\begin{align}
p\Big( \vp_i^{(\alpha)} \Big| \{c_{ij}^{(\alpha-1)}\}\Big) = p\Big(\vp_i^{(\alpha)} \Big) = 1/(2\pi)^{N}
\end{align}
or a pair of $\left(\vp_i^{(\alpha)},\vp_j^{(\alpha)}\right)$ is given by
\begin{align}
p\Big( \vp_i^{(\alpha)},\vp_j^{(\alpha)} \Big| \{c_{ij}^{(\alpha-1)}\}\Big) = p\Big(\vp_i^{(\alpha)},\vp_j^{(\alpha)} \Big) =  1/(2\pi)^{N}.
\end{align}

Next, we compute the expected amplitude factor using a Bayesian approach as in the single-device protocol. Here, we simply present the analytic computation of expectation values, since we obtain the same moments as the single-device case: for the expectations value of the amplitude factor $\log \big[ S_{(i,j),\ket{00}}^{(\alpha)} \big]$ we get
\begin{align}
    \mu_{(i,j),\alpha,\ket{00}} =& \mathbb{E}\left[\log \big[ S_{(i,j),\ket{00}}^{(\alpha)} \big] \Big| \ket{0}, \{c_{ij}^{(\alpha-1)}\} \right] \\
    =& \frac{1}{(2\pi)^{N}} \int_0^{2\pi} \cdots \int_0^{2\pi} \log \big[1+\cos(\vp_i^{(k)}+\vp_j^{(k)})\big] \nonumber \\
    &\times \left( \sum_{l,p}|c_{lp}^{(\alpha-1)}|^2 \big[1+\cos(\vp_l^{(k)}+\vp_p^{(k)})\big] \right) d\vp_1^{(\alpha)}\dots d\vp_{N}^{(\alpha)} \\
    =& |c_{ij}^{(\alpha-1)}|^2 - \log(2).
\end{align}
We have $ \mu_{(i,j),\alpha,\ket{11}} =  \mu_{(i,j),\alpha,\ket{00}}$ due to symmetries. Hence, the ``00" and ``11" bit string replicate the behavior of the single device case and occur here with total probability $1/2$. The ``10" bit string corresponds to a projection onto the antisymmetric space which sets the identical eigenstate amplitudes $|c_{dd}|^2$ to zero. 

It remains to analyze the behavior of the circuit for the ``01" bit string. 
First, we compute $\mu_{(d,d),\alpha,\ket{01}}$ where we do not need to integrate because $S_{(d,d),\ket{01}}^{(\alpha)}$ is independent of the $\vp_i^{(\alpha)}$:
\begin{align}
    \mu_{(d,d),\alpha,\ket{01}} =& \mathbb{E}\left[\log \big[ S_{(d,d),\ket{01}}^{(\alpha)} \big] \Big| \ket{0}, \{c_{ij}^{(\alpha-1)}\} \right] \\
    =& \mathbb{E}\left[\log \big[1+\cos(\vp_d^{(\alpha)}-\vp_d^{(\alpha)})\big] \Big| \ket{0}, \{c_{ij}^{(\alpha-1)}\} \right] \\ 
    =& \log(2). 
\end{align}
Next, we compute $\mu_{(d,j),\alpha,\ket{01}}$ where $d\neq j$. First, we write the biased probability density function for the $\vp_i^{(\alpha)}$, which is given as
\begin{align}
    p\Big(\vp_i^{(\alpha)} \Big| \ket{01}, \{c_{ij}^{(\alpha-1)}\} \Big) &=
    \frac{ p \Big(\ket{01} \Big| \{\vp_i^{(\alpha)}\} , \{c_{ij}^{(\alpha-1)}\} \Big) \times p\Big(\vp_i^{(\alpha)} \Big| \{c_{ij}^{(\alpha-1)}\}\Big)} {p\Big(\ket{01} \Big| \{c_{ij}^{(\alpha-1)}\} \Big)} \\
    &= \frac{1}{(2\pi)^{N}} \frac{1}{1 + \sum_l |c_{ll}^{(\alpha-1)}|^2} \sum_{i,j}|c_{ij}^{(\alpha-1)}|^2 \big[1+\cos(\vp_i^{(\alpha)}-\vp_j^{(\alpha)})\big].
\end{align}
Then, the expectation value of $\log \big[ S_{(d,j),\ket{01}}^{(\alpha)} \big]$ is
\begin{align}
    \mu_{(d,j),\alpha,\ket{01}} 
    =& \mathbb{E}\left[\log \big[ S_{(d,j),\ket{01}}^{(\alpha)} \big] \Big| \ket{0}, \{c_{ij}^{(\alpha-1)}\} \right] \\
    =& \frac{1}{1 + \sum_l |c_{ll}^{(\alpha-1)}|^2}\frac{1}{(2\pi)^{N}} \int_0^{2\pi} \cdots \int_0^{2\pi} \log \big[1+\cos(\vp_d^{(\alpha)}-\vp_j^{(\alpha)})\big] \nonumber \\
    &\times \sum_{l,p}|c_{lp}^{(\alpha-1)}|^2 \big[1+\cos(\vp_l^{(\alpha)}-\vp_p^{(\alpha)})\big] d\vp_1^{(\alpha)}\dots d\vp_{N}^{(\alpha)} \\
    =& \frac{1}{1 + \sum_l |c_{ll}^{(\alpha-1)}|^2} \int_0^{2\pi} \cdots \int_0^{2\pi} \Biggl\{ \log \big[\dots\big]_{dj} \sum_{a}|c_{aa}^{(\alpha-1)}|^2 \big[\dots\big]_{aa} +
     \log \big[\dots\big]_{dj} |c_{dj}^{(\alpha-1)}|^2 \big[\dots\big]_{dj} \nonumber\\
     &+ \log \big[\dots\big]_{dj} |c_{jd}^{(\alpha-1)}|^2 \big[\dots\big]_{dj} 
    + \log \big[\dots\big]_{dj} \sum_{(a,b)\in R} |c_{ab}^{(\alpha-1)}|^2 \big[\dots\big]_{ab} \Biggr\}\; d\vp_1^{(\alpha)}\dots d\vp_{N}^{(\alpha)}  \\
    =& \frac{1}{1 + \sum_l |c_{ll}^{(\alpha-1)}|^2} \Biggl\{ -2\log(2) \sum_{a}|c_{aa}^{(\alpha-1)}|^2 + (1-\log(2))|c_{dj}^{(\alpha-1)}|^2 \nonumber\\ 
    &+ (1-\log(2)) |c_{jd}^{(\alpha-1)}|^2 - \log(2) \sum_{(a,b)\in R} |c_{ab}^{(\alpha-1)}|^2 \Biggr\} \\
    =& -\log(2) + \frac{|c_{dj}|^2 + |c_{jd}|^2}{1 + \sum_l |c_{ll}^{(\alpha-1)}|^2}, \label{eq:mudj}
\end{align}
where $R^c:=\bigcup_a\{(a,a)\} \cup \{(d,j)\}\cup \{(j,d)\}$ and we use the shorthand $\big[\dots\big]_{lp}:= \big[1+\cos(\vp_l^{(\alpha)}-\vp_p^{(\alpha)})\big]$. If we had used a uniform distribution for the $\vp_i^{(\alpha)}$, we would have obtained $-\log(2)$ instead, i.e.~the second term in Eq.~\ref{eq:mudj} is a correction which goes to zero if the non-dominant amplitudes $|c_{dj}|^2$ and $|c_{jd}|^2$ vanish. Note that in the broadcasting setup $|c_{jd}|^2\rightarrow 0$ as we assume that Alice has a near-perfect approximation to the ground state.

\subsection{Average relative suppression with entanglement sharing} \label{ssec:AB_suppfactor}
Akin to the single device case, we compute the average suppression per round of the circuit in the near-convergent regime where $|c_{dd}^{(\alpha-1)}|^2 \rightarrow_\alpha 1$ and $|c_{ds}^{(\alpha-1)}|^2 \rightarrow_\alpha 0$. The expected suppression factor per round is given as (cf.~\ref{eq:variancevanishes})
\begin{align}
     \exp \left( \mu_{(d,s),\alpha,\ket{01}} - \mu_{(d,d),\alpha,\ket{01}} \right) 
     = \exp \left(-2\log(2) \right) = 1/4 \label{eq:AB_1/4}
\end{align}
in the limit $m\rightarrow\infty$ for a measured ``01" bit string. Now combining the computed expectation values, for the ``00" and ``11" measurements, the suppression factor converges to $1/e$ in the near-convergent regime. Hence, the total suppression factor weighted by their probability of occurrence, given that we always project to the symmetric subspace, is 
\begin{align}
    \frac{1}{e^{1/2}} \frac{1}{4^{1/2}} = \frac{1}{2\sqrt{e}} \approx 0.30
\end{align}
and smaller than the value for the single device case where we had $1/e \approx 0.37$.

\subsection{Overhead due to unsuccessful projection} \label{ssec:post}

For noisy quantum devices with limited coherence time, a successful algorithmic strategy is to improve the figure of merit of the quantum algorithm at the expense of more repetitions of the protocol. In this spirit, we use heralding through the measured bit string to assess whether the protocol needs to be restarted. Here, we estimate the algorithmic overhead due to a ``10" bit string being observed, which requires the protocol to be restarted. The probability of observing ``10" was stated in Eq.~\ref{eq:AB_p10} and depends on the diagonal probabilities $\sum_i |c_{ii}^{(\alpha-1)}|^2$. We introduce the notation $\sum_{i,j}|c_{ij}|^2 = ||\vec{C}||_2^2= ||\vec{C}_\mathrm{d}||^2_2+||\vec{C}_\mathrm{od}||^2_2$=1, where
\begin{align}
    ||\vec{C}_\mathrm{d}||^2_2 = \sum_i |c_{ii}|^2
\end{align}
is the norm of the diagonal elements.
Let us focus on proving the conditions for convergence to the diagonal; i.e., to $c_{i,j}=0$ for all $i\neq j$.

To a first approximation, we consider the effect of an observed bit string on $||\vec{C}_\mathrm{d}||^2$. For the bit strings ``00" and ``11" we take a conservative approach in that the diagonal norm remains indifferent. For a ``10" the relative suppression of the off-diagonal $c_{i,j}$ is $1/4$ in the near-convergent limit. We consider a regime where the dependence of the $S_{(i,j),\ket{b}}^{(k)}$ is sufficiently weak such that the Central Limit Theorem holds, so we may compute an average suppression factor by comparing $\mu_{(d,j),\alpha,\ket{01}}$ with $\mu_{(d,d),\alpha,\ket{01}}$ as we do in Sec.~\ref{ssec:AB_suppfactor}. We bound $\mu_{(d,j),\alpha,\ket{01}} < -\log(2) + 1/2$ as $|c_{dj}|^2<1/2$ and $|c_{jd}|^2=0$ in the broadcasting scenario. Then the average suppression factor for an observed ``01" is smaller (i.e.~better) than 
\begin{align}
    \exp(-2\log(2)+1/2) = \sqrt{e}/4 \approx 0.41 < 1/2.
\end{align}

Hence, we have that the new expected amplitudes after one iteration transforms on average as least as favorable as 
\begin{align}
    \mathbbm{E}\left[||\vec{C^{(k)}}_\mathrm{d}||^2_2\right] = \left\{
    \begin{array}{ccc}
    \displaystyle ||\vec{C}^{(k-1)}_\mathrm{d}||^2_2 & \mbox{with expected prob.} & p\Big(\ket{00}\vee\ket{11} \Big| \{c_{ij}^{(\alpha-1)}\}\Big)=1/2,\\[10pt]
    \displaystyle \frac{2||\vec{C}^{(k-1)}_\mathrm{d}||^2_2}{1+||\vec{C}^{(k-1)}_\mathrm{d}||^2_2}&\mbox{with expected prob.}&p\Big(\ket{01} \Big| \{c_{ij}^{(\alpha-1)}\}\Big) = \frac{1}{4}\left(1+||\vec{C}_\mathrm{d}||^2_2\right),\\[10pt]
    \displaystyle 0&\mbox{with expected prob.}& p\Big(\ket{10} \Big| \{c_{ij}^{(\alpha-1)}\}\Big)=\frac{1}{4}\left(1-||\vec{C}_\mathrm{d}||^2_2\right).
    \end{array}
    \right.
\end{align}
Let $x^{(k)} = \mathbbm{E}\left[||\vec{C}^{(k)}_\mathrm{d}||^2_2\right]$. In this simplified model, we need only study under which conditions $\lim_{k\rightarrow \infty}x^{(k)}=1$.
Solving the difference equation for $x^{(k)}$ yields
\begin{align}
    x^{(k)} = \frac{2^kx^{(0)}}{1+x^{(0)}(2^k-1)}.
\end{align}

The probability to observe the ``01" bit string $p$ times, conditioned that we do not count outcomes ``00" or ``11" (since their ability to increase $||\vec{C}^{(k)}_\mathrm{d}||^2_2$ is limited and will be ignored here) is given by
\begin{align}
    \prod_{k=1}^{p}\left(\frac{1}{2} + \frac{x^{(k)}}{2}\right) = \frac{1+(2^{p+1}-1)x^{(0)}}{2^p(x^{(0)}+1)},
\end{align}
which tends asymptotically to $2x^{(0)}/(1+x^{(0)})$.

Hence, we expect that the probability to see a projection to the antisymmetric space (bit string ``10") at some iteration will be given by
\begin{align}
1 - \frac{2x^{(0)}}{1+x^{(0)}} = 
    \frac{1-||\vec{C}^{(0)}_\mathrm{d}||^2_2}{1+||\vec{C}^{(0)}_\mathrm{d}||^2_2}
\end{align}
and therefore, for an initial ground state overlap $\gamma=|c^{(0)}_{B,0}|$ with $||\vec{C}^{(0)}_\mathrm{d}||^2_2 = \gamma^2 + \sum_{i>0} |c_{ii}|^2$ (in the broadcasting setup) is the increase of the total algorithmic cost due to restarts given as $\mathcal{O}(1/\gamma^{2})$.

\clearpage
\section{Quantum broadcasting: 3-auxiliary qubit variant ($V=H \otimes H$)} \label{sec:threeancillas}
We discuss a variant of the protocol where the two auxiliary qubits controlling the dynamics are projected to the symmetric subspace. Then, these two qubits are measured in the $\sigma^x$-basis by applying Hadamard gates before a computational basis measurement. The circuit for one iteration of the protocol is shown in Fig.~\ref{fig:2Hadamards} and it is obtained by choosing the operator $V=H\otimes H$ for in the circuit diagram from the main text that includes the controlled SWAP. This operator $V$ reads
\begin{align}
    V = H\otimes H = \frac{1}{2}\begin{pmatrix}
    1 & 1 & 1 & 1\\
    1 & -1 & 1 & 1\\
    1 & 1 & -1 & 1\\
    1 & -1 & -1 & 1
\end{pmatrix},
\end{align}

\begin{figure}[htbp]
    \centering
    \begin{tikzpicture}
    \node[scale=1] {
        \begin{quantikz}
            \lstick{\ket{0}} &  \gate{H} &  \qw &  \qw  & \ctrl{2} &  \gate{H} &  \meter{} \arrow[r] & \rstick{$\ket{0}$} \\
             \lstick{\ket{0}} &  \gate{H}  &  \ctrl{2} &  \qw &  \swap{1} & \gate{H} &  \meter{} & \\
             \lstick{\ket{0}} &  \gate{H}  & \qw  &  \ctrl{2} & \targX{}  & \gate{H}  & \meter{}& \\
            \lstick[wires=2]{$\ket{\Psi^{(k-1)}}$}  & \qw &  \gate{U(\tau_k)} &   \qw &  \qw &  \qw &  \qw\rstick[wires=2]{$\ket{\Psi^{(k)}}$} & \\
              &  \qw &  \qw &   \gate{U(\tau_k)} &  \qw &  \qw &  \qw  &
        \end{quantikz}
        };
    \end{tikzpicture}%
    \caption{Quantum broadcasting circuit with an additional auxiliary qubit for the projection to the symmetric subspace. In the protocol we postselect on observing a $\ket{0}$ on the first helper qubit.}
    \label{fig:2Hadamards}
\end{figure}

\subsection{Quantum circuit and process map}

We begin with the action of the circuit (which includes for this variant now the controlled SWAP gate) for a general $V$. We denote $\ket{\psi'}:=U\ket{\psi}$. Again, we begin considering the case where $\ket{\Psi^{(0)}} = \ket{\psi}\ket{\psi}$. Before the computational basis measurements the circuit is preparing the global state
\begin{align}
 & \frac{\ket{0}}{4} \Big[ 2 V\ket{00} \ket{\psi \psi} + V\ket{01} \big( \ket{\psi \psi'} + \ket{\psi' \psi} \big) +  V\ket{10} \big( \ket{\psi \psi'} + \ket{\psi' \psi} \big) + 2 V\ket{11} \ket{\psi' \psi'} \Big] \nonumber \\ 
 +&  \frac{\ket{1}}{4} \Big[ V\ket{01} \big( \ket{\psi \psi'} - \ket{\psi' \psi} \big) +  V\ket{10} \big( \ket{\psi' \psi} - \ket{\psi \psi'} \big) \Big].
\end{align}
For a choice of $V=H\otimes H$ we obtain
\begin{align}
    \frac{\ket{0}}{4} \Big[ &\ket{00} ( \ket{\psi \psi} +  \ket{\psi \psi'} + \ket{\psi' \psi} +\ket{\psi' \psi'}) ) 
    + \ket{01} ( \ket{\psi \psi}    -\ket{\psi' \psi'}) ) \nonumber\\ 
    + &\ket{10} ( \ket{\psi \psi}    -\ket{\psi' \psi'}) )  
    + \ket{11} ( \ket{\psi \psi} -  \ket{\psi \psi'} - \ket{\psi' \psi} +\ket{\psi' \psi'}) ) \Big] \nonumber\\
    +  \frac{\ket{1}}{4} \Big[ &\ket{01} ( \ket{\psi' \psi}    - \ket{\psi \psi'}) ) +\ket{10} ( \ket{\psi \psi'}    - \ket{\psi' \psi}) ) \Big].
\end{align}
We observe that in this variant, again, the probability to measure a $\ket{1}$ on the first auxiliary qubit approaches zero as the algorithm converges towards an eigenstate $\ket{\psi} \approx \ket{\phi_j}$. Now we express the state in the eigenbasis of $H$:
\begin{align}
    \frac{\ket{0}}{4} \Big[ &\ket{00} \sum_{i,j} c_{ij} (1 + e^{i\vp_i} + e^{i\vp_j} + e^{i(\vp_i+\vp_j)})\ket{\phi_i \phi_j} 
    + \ket{01} \sum_{i,j} c_{ij} (1 - e^{i(\vp_i+\vp_j)})\ket{\phi_i \phi_j} \nonumber\\ 
    + &\ket{10} \sum_{i,j} c_{ij} (1 - e^{i(\vp_i+\vp_j)})\ket{\phi_i \phi_j} )  
    + \ket{11} \sum_{i,j} c_{ij} (1 - e^{i\vp_i} - e^{i\vp_j} + e^{i(\vp_i+\vp_j)})\ket{\phi_i \phi_j}  \Big]\nonumber\\
    \frac{\ket{1}}{4} \Big[ & \ket{01} \sum_{i,j} c_{ij} (e^{i\vp_i} - e^{i\vp_j})\ket{\phi_i \phi_j} + \ket{10} \sum_{i,j} c_{ij} (e^{i\vp_j} - e^{i\vp_i})\ket{\phi_i \phi_j} \Big].
\end{align}
The process map for the quantum broadcasting protocol is given by  
\begin{align}
|c_{ij}^{(k-1)}|^2\mapsto & |c_{ij}^{(k)}|^2 \\
=& |c_{ij}^{(k-1)}|^2\times \frac{S_{ij}^{(k)}}{N^{(k)}} \\
=& |c_{ij}^{(k-1)}|^2\times \left\{
\begin{array}{ccc}
      \big[1+\cos(\vp_i^{(k)})\big]\big[1+\cos(\vp_j^{(k)})\big]/\big( 4 N_{\ket{000}}^{(k)} \big) &  \mbox{with} & \Pr{}_{\ket{000}}^{(k)},\\[10pt]
     \big[1-\cos(\vp_i^{(k)}+\vp_j^{(k)})\big]/\big( 8 N_{\ket{001}}^{(k)} \big) &  \mbox{with} & \Pr{}_{\ket{001}}^{(k)},\\[10pt]
     \big[1-\cos(\vp_i^{(k)}+\vp_j^{(k)})\big]/\big( 8 N_{\ket{010}}^{(k)} \big) &  \mbox{with} & \Pr{}_{\ket{010}}^{(k)},\\[10pt]
     \big[1-\cos(\vp_i^{(k)})\big]\big[1-\cos(\vp_j^{(k)})\big]/\big( 4 N_{\ket{011}}^{(k)} \big) &  \mbox{with} & \Pr{}_{\ket{011}}^{(k)},
\end{array}
\right.    
\end{align}
where we only consider the classical bit strings with a ``0'' on the first auxiliary qubit, since in the protocol we postselect on this measurement outcome, which corresponds to a projection to the symmetric subspace. Going forward, we again absorb the factor of $1/4$ or $1/8$, respectively, into the normalization.
The probabilities for measuring the different bit strings are
\begin{align} 
    \Pr{}_{\ket{000}}^{(\alpha)} &= \frac{1}{4}\sum_{i,j}|c_{ij}^{(\alpha-1)}|^2 \big[1+\cos(\vp_i^{(k)})\big]\big[1+\cos(\vp_j^{(k)})\big],\\
    \Pr{}_{\ket{001}}^{(\alpha)} &= \Pr{}_{\ket{010}}^{(\alpha)} = \frac{1}{8}\sum_{i,j}|c_{ij}^{(\alpha-1)}|^2 \big[1-\cos(\vp_i^{(k)}+\vp_j^{(k)})\big],\\
    \Pr{}_{\ket{011}}^{(\alpha)} &= \frac{1}{4}\sum_{i,j}|c_{ij}^{(\alpha-1)}|^2 \big[1-\cos(\vp_i^{(k)})\big]\big[1-\cos(\vp_j^{(k)})\big]
\end{align}
and, for completeness, we also include the antisymmetric cases:
\begin{align} 
    \Pr{}_{\ket{100}}^{(\alpha)} &= \Pr{}_{\ket{111}}^{(\alpha)} = 0, \\
    \Pr{}_{\ket{101}}^{(\alpha)} &= \Pr{}_{\ket{110}}^{(\alpha)} = \frac{1}{8}\sum_{i,j}|c_{ij}^{(\alpha-1)}|^2 \big[1-\cos(\vp_i^{(k)}-\vp_j^{(k)})\big].
\end{align}
We can indeed check that the probabilities sum to one $\sum_{b \in \{0,1\}^3} \Pr{}_{\ket{b}}^{(\alpha)}$ = 1.
In the following, we write the probabilities as a conditional probability $\Pr{}_{\ket{000}}^{(\alpha)} = p\Big(\ket{000} \Big| \vp_i^{(\alpha)}, \{c_{ij}^{(\alpha-1)}\}\Big)$, etc., as the probability depends on the coefficients and the product of the evolution time with the eigenenergies.

\subsection{Bayesian approach for the expected amplitude factor} \label{sec:AB_Bayes}

We compute the expected amplitude factor using a Bayesian approach as for the single device protocol. Here, we restrict the analytic computation to expectation values and confirm numerically that variances are finite. We require the distribution of the $\vp_i^{(\alpha)}$ for the calculation of the first moment. 
Analogously to the previous sections, we list the terms that we need to plug into the Bayesian formula
\begin{align}
    & p\Big(\ket{000} \Big| \{c_{ij}^{(\alpha-1)}\}\Big) = p\Big(\ket{011} \Big| \{c_{ij}^{(\alpha-1)}\}\Big) = \frac{1}{4} + \frac{1}{8}\sum_i |c_{ii}^{(\alpha-1)}|^2, \label{eq:prob000} \\
    & p\Big(\ket{001} \Big| \{c_{ij}^{(\alpha-1)}\}\Big) = p\Big(\ket{010} \Big| \{c_{ij}^{(\alpha-1)}\}\Big) = \frac{1}{8}.
\end{align}
From the Bayesian rule we have
\begin{align}
    p\Big(\vp_i^{(\alpha)} \Big| \ket{001}, \{c_{ij}^{(\alpha-1)}\} \Big) &=
    \frac{ p \Big(\ket{001} \Big| \vp_i^{(\alpha)} , \{c_{ij}^{(\alpha-1)}\} \Big) \times p\Big(\vp_i^{(\alpha)} \Big| \{c_{ij}^{(\alpha-1)}\}\Big)} {p\Big(\ket{001} \Big| \{c_{ij}^{(\alpha-1)}\} \Big)} \\
    &= \frac{1}{(2\pi)^{N}} \sum_{i,j}|c_{ij}^{(\alpha-1)}|^2 \big[1-\cos(\vp_i^{(k)}+\vp_j^{(k)})\big]
\end{align}
and we analyze the expectation value of the amplitude factor for the $\ket{001}$ (and $\ket{010}$) measurement. We can compute $\mu_{(i,j),\alpha,\ket{001}} = \mu_{(i,j),\alpha,\ket{010}}$ as
\begin{align}
    \mu_{(i,j),\alpha,\ket{001}} =& \mathbb{E}\left[\log \big[ S_{(i,j),\ket{001}}^{(\alpha)} \big] \Big| \ket{001}, \{c_{ij}^{(\alpha-1)}\} \right] \\
    =& \frac{1}{(2\pi)^{N}} \int_0^{2\pi} \cdots \int_0^{2\pi} \log \big[1-\cos(\vp_i^{(k)}+\vp_j^{(k)})\big] \nonumber\\
    &\times \left( \sum_{l,p}|c_{lp}^{(\alpha-1)}|^2 \big[1-\cos(\vp_i^{(k)}+\vp_j^{(k)})\big] \right) d\vp_1^{(\alpha)}\dots d\vp_{N}^{(\alpha)} \\
    =& |c_{ij}^{(\alpha-1)}|^2 - \log(2),
\end{align}
which shows the same suppression behavior as in the single device case.

Now we consider the outcome of a ``000" bit string. From the Bayesian rule we obtain
\begin{align}
    p\Big(\vp_i^{(\alpha)} \Big| \ket{000}, \{c_{ij}^{(\alpha-1)}\} \Big) &=
    \frac{ p \Big(\ket{000} \Big| \vp_i^{(\alpha)} , \{c_{ij}^{(\alpha-1)}\} \Big) \times p\Big(\vp_i^{(\alpha)} \Big| \{c_{ij}^{(\alpha-1)}\}\Big)} {p\Big(\ket{000} \Big| \{c_{ij}^{(\alpha-1)}\} \Big)} \\
    &= \frac{1}{1 + \frac{1}{2}\sum_a |c_{aa}^{(\alpha-1)}|^2} \frac{1}{(2\pi)^{N}} \sum_{i,j}|c_{ij}^{(\alpha-1)}|^2 \big[1+\cos(\vp_i^{(k)})\big]\big[1+\cos(\vp_j^{(k)})\big].
\end{align}
Next, we can calculate $\mu_{(i,j),\alpha,\ket{000}}$:
\begin{align}
    & \mu_{(i,j),\alpha,\ket{000}} \\
    =& \mathbb{E}\left[\log \big[ S_{(i,j),\ket{000}}^{(\alpha)} \big] \Big| \ket{000}, \{c_{ij}^{(\alpha-1)}\} \right] \\
    =&  \frac{1}{1 + \frac{1}{2}\sum_a |c_{aa}^{(\alpha-1)}|^2} \frac{1}{(2\pi)^{N}} \int_0^{2\pi} \cdots \int_0^{2\pi} \log \left( \big[1+\cos(\vp_i^{(k)})\big]\big[1+\cos(\vp_j^{(k)})\big] \right) \nonumber\\
    &\times \Biggl( \sum_{l,p}|c_{lp}^{(\alpha-1)}|^2 \big[1+\cos(\vp_l^{(k)})\big]\big[1+\cos(\vp_p^{(k)})\big] \Biggr) d\vp_1^{(\alpha)}\dots d\vp_{N}^{(\alpha)} \\
    =&  \frac{(2\pi)^{-N}}{1 + \frac{1}{2}\sum_a |c_{aa}^{(\alpha-1)}|^2}  \int_0^{2\pi} \cdots \int_0^{2\pi} \Biggl\{ \log \big[\dots\big]_{i} \Biggl( \sum_{l,p}|c_{lp}^{(\alpha-1)}|^2 \big[\dots\big]_{lp} \Biggr) + i\leftrightarrow j \Biggr\} d\vp_1^{(\alpha)}\dots d\vp_{N}^{(\alpha)}   \\
    =& \frac{(2\pi)^{-N}}{1 + \frac{1}{2}\sum_a |c_{aa}^{(\alpha-1)}|^2}  \int_0^{2\pi} \cdots \int_0^{2\pi} \Biggl\{ \log \big[\dots\big]_{i} \Biggl( |c_{ii}^{(\alpha-1)}|^2 \big[\dots\big]_{ii} + \sum_{a \neq i} \big(|c_{ia}^{(\alpha-1)}|^2 + |c_{ai}^{(\alpha-1)}|^2\big) \big[\dots\big]_{ia} \nonumber\\
    &+  \sum_{a \neq i} |c_{aa}^{(\alpha-1)}|^2 \big[\dots\big]_{aa} +  \sum_{a \neq b \neq i \neq a} |c_{ab}^{(\alpha-1)}|^2 \big[\dots\big]_{ab}  \Biggr) + i\leftrightarrow j \Biggr\} d\vp_1^{(\alpha)}\dots d\vp_{N}^{(\alpha)}   \\
    =& \frac{1}{1 + \frac{1}{2}\sum_a |c_{aa}^{(\alpha-1)}|^2}  \Biggl\{ \Biggl( |c_{ii}^{(\alpha-1)}|^2 \biggl[\frac{7}{4} - \frac{3}{2}\log(2)\biggr] + \sum_{a \neq i} \big(|c_{ia}^{(\alpha-1)}|^2 + |c_{ai}^{(\alpha-1)}|^2\big) \big[1-\log(2)\big] \nonumber\\
    &+  \sum_{a \neq i} |c_{aa}^{(\alpha-1)}|^2 \biggl[-\frac{3}{2}\log(2)\biggr] +  \sum_{a \neq b \neq i \neq a} |c_{ab}^{(\alpha-1)}|^2 \big[-\log(2)\big]  \Biggr) + i\leftrightarrow j \Biggr\}  \\
    =& \frac{1}{1 + \frac{1}{2}\sum_a |c_{aa}^{(\alpha-1)}|^2}  \Biggl\{ \Biggl( -\log(2) + |c_{ii}^{(\alpha-1)}|^2 \biggl[\frac{7}{4} - \frac{1}{2}\log(2)\biggr] + \sum_{a \neq i} \big(|c_{ia}^{(\alpha-1)}|^2 + |c_{ai}^{(\alpha-1)}|^2\big) \nonumber\\
    &+  \sum_{a \neq i} |c_{aa}^{(\alpha-1)}|^2 \biggl[-\frac{1}{2}\log(2)\biggr] \Biggr) + i\leftrightarrow j \Biggr\}  \\
    =& \frac{1}{1 + \frac{1}{2}\sum_a |c_{aa}^{(\alpha-1)}|^2}  \Biggl\{ \Biggl( -\log(2) \biggl(1+\frac{1}{2}\sum_a |c_{aa}^{(\alpha-1)}|^2\biggr) + \frac{7}{4}|c_{ii}^{(\alpha-1)}|^2  + \sum_{a \neq i} \big(|c_{ia}^{(\alpha-1)}|^2 + |c_{ai}^{(\alpha-1)}|^2\big) \Biggr) + i\leftrightarrow j \Biggr\}  \\
    =& -2\log(2) + \frac{1}{1 + \frac{1}{2}\sum_a |c_{aa}^{(\alpha-1)}|^2} \biggl[ \frac{7}{4}|c_{ii}^{(\alpha-1)}|^2 + \frac{7}{4}|c_{jj}^{(\alpha-1)}|^2 + \sum_{a \neq i} \big(|c_{ia}^{(\alpha-1)}|^2 + |c_{ai}^{(\alpha-1)}|^2\big) + \sum_{a \neq j} \big(|c_{ja}^{(\alpha-1)}|^2 + |c_{aj}^{(\alpha-1)}|^2\big) \biggr]
\end{align}
where we used the shorthands $\big[\dots\big]_{l}:= \big[1+\cos(\vp_l^{(k)})\big]$ and $\big[\dots\big]_{lp}:= \big[1+\cos(\vp_l^{(k)})\big] \big[1+\cos(\vp_p^{(k)})\big]$. For the near-convergent regime where $|c_{dd}|^2\rightarrow1$, we obtain
\begin{align}
    \mu_{(d,d),\alpha,\ket{000}} = -2\log(2) + \frac{7}{3}
\end{align}
and
\begin{align}
    \mu_{(d,s),\alpha,\ket{000}} = -2\log(2) + \frac{7}{6},
\end{align}
respectively. Due to symmetry of the trigonometric functions we have $\mu_{(i,j),\alpha,\ket{000}} = \mu_{(i,j),\alpha,\ket{011}}$.

\subsection{Average relative suppression with entanglement sharing} \label{ssec:AB_relsupp}

We computed the average amplitude factors for $\log \big[ S_{(d,j),\ket{b}}^{(\alpha)} \big]$. The variances are not analytically evaluated as they are not required in order to compute the average relative suppression of undesired amplitudes. Numerical computation of the variances shows that they are indeed finite which is necessary for the central limit theorem to be applicable.
Finally, in the regime considered in Sec.~\ref{sec:AB_Bayes}, we evaluate the average relative suppression per round for the ``000" (or ``011") measurement as
\begin{align}
     \exp \left( \mu_{(d,s),\alpha,\ket{000}} - \mu_{(d,d),\alpha,\ket{000}} \right) 
     = \exp \left( -7/6 \right)
\end{align}
in the limit $m\rightarrow\infty$. For the ``001" and ``010" measurements, by analogy to the single device case, the suppression factor per round converges to $\exp (-1)$.
Taking into account the probabilities for each of the bit strings to occur in the near-convergent regime, the overall suppression factor per round is
\begin{align}
    \exp(-7/6)^{6/8} \exp(-1)^{2/8} = \exp(-9/8),
\end{align}
improving over $\exp(-1)$ for the single device protocol.

\clearpage
\section{Protocol for projection into joint product basis}

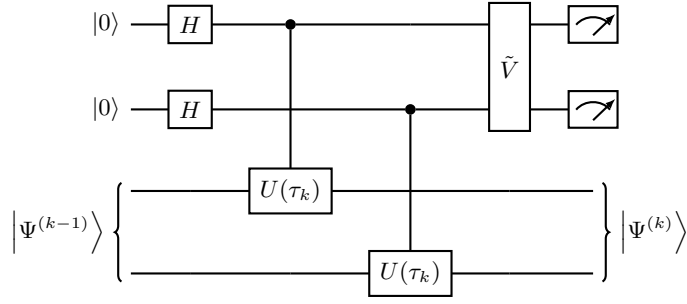
\begin{figure}[h]
    \centering
    \begin{tikzpicture}
    \node[scale=1] {
        \begin{quantikz}
             \lstick{\ket{0}} &  \gate{H}  &  \ctrl{2} &  \qw &  \gate[wires=2]{\tilde V} &  \meter{} & \\
             \lstick{\ket{0}} &  \gate{H}  & \qw  &  \ctrl{2}   &  & \meter{}& \\
            \lstick[wires=2]{$\ket{\Psi^{(k-1)}}$}  & \qw &  \gate{U(\tau_k)} &   \qw &  \qw &   \qw\rstick[wires=2]{$\ket{\Psi^{(k)}}$} & \\
              &  \qw &  \qw &   \gate{U(\tau_k)} &    \qw &  \qw  &
        \end{quantikz}
        };
    \end{tikzpicture}%
    \caption{Circuit for the variant for a projection into a joint product basis of two devices.}
    \label{fig:circuitproductbasis}
\end{figure}

Here, we consider a different matrix $\tilde V$ which allows to project two quantum state on Alice and Bob respectively to their common product eigenbasis. The structure of this state is $\ket{\Psi} = \sum_{i} c_{AB,ii} \ket{\phi_i}\ket{\phi_i}$, i.e., a Schmidt decomposition in the eigenbasis. Such states are central in quantum information processing and they might be of wider use for other quantum algorithms. For the operator $\tilde V$, here we shall use
\begin{align}
    \tilde V = \frac{1}{\sqrt{2}} \begin{pmatrix}
    \sqrt{2} & 0 & 0 & 0 \\
    0 & 1 & 1 & 0 \\
    0 & 1 & -1 & 0 \\
    0 & 0 & 0 & \sqrt{2} 
    \end{pmatrix}
\end{align}
and repeat the key steps of the calculation. We insert the new choice for $\tilde V$ in Eq.~\ref{eq:withV} and obtain
\begin{align}
    & \frac{1}{2\sqrt{2}} \Big[ \sqrt{2}\ket{00} \ket{\psi \psi}   + \ket{01}  \big( \ket{\psi \psi'} + \ket{\psi' \psi} \big)  +  \sqrt{2} \ket{11} \ket{\psi' \psi'} + \ket{10}  \big( \ket{\psi' \psi} - \ket{\psi \psi'} \big)  \Big] \\
    =& \frac{1}{2\sqrt{2}} \Big[ \sqrt{2}\ket{00} \sum_{i,j} c_{ij} \ket{\phi_i \phi_j}
    + \ket{01} \sum_{i,j} c_{ij} (e^{i\vp_i} + e^{i\vp_j})\ket{\phi_i \phi_j}  \nonumber \\ 
    &+ \sqrt{2} \ket{11} \sum_{i,j} c_{ij} e^{i(\vp_i + \vp_j)} \ket{\phi_i \phi_j} ) + \ket{10} \sum_{i,j} c_{ij} (e^{i\vp_i} - e^{i\vp_j})\ket{\phi_i \phi_j}  \Big].
\end{align}
The process map is then given as
\begin{align}
    |c_{ij}^{(k-1)}|^2\mapsto & |c_{ij}^{(k)}|^2 \\
    =& |c_{ij}^{(k-1)}|^2 \times \frac{S_{(i,j),\ket{b}}^{(k)}}{N^{(k)}_{\ket{b}}} \\
    =& |c_{ij}^{(k-1)}|^2\times \left\{
    \begin{array}{ccc}
        1/\big( 4 N_{\ket{00}}^{(k)} \big) &  \mbox{with} & \Pr{}_{\ket{00}}^{(k)},\\[10pt]
        \big[1+\cos(\vp_i^{(k)}-\vp_j^{(k)})\big]/\big( 4 N_{\ket{01}}^{(k)} \big) &  \mbox{with} & \Pr{}_{\ket{001}}^{(k)},\\[10pt]
        1/\big( 4 N_{\ket{11}}^{(k)} \big) &  \mbox{with} & \Pr{}_{\ket{11}}^{(k)},\\[10pt]
        \big[1-\cos(\vp_i^{(k)}-\vp_j^{(k)})\big]/\big( 4 N_{\ket{10}}^{(k)} \big) &  \mbox{with} & \Pr{}_{\ket{10}}^{(k)}.
    \end{array}
    \right.    
\end{align}
The relevant probabilities are
\begin{align} 
    \Pr{}_{\ket{00}}^{(\alpha)} &=  \Pr{}_{\ket{11}}^{(\alpha)} = \frac{1}{4},\\
    \Pr{}_{\ket{01}}^{(\alpha)} &= \frac{1}{4}\sum_{i,j}|c_{ij}^{(\alpha-1)}|^2  \big[1+\cos(\vp_i^{(k)}-\vp_j^{(k)})\big],\\
    \Pr{}_{\ket{10}}^{(\alpha)} &= \frac{1}{4}\sum_{i,j}|c_{ij}^{(\alpha-1)}|^2  \big[1-\cos(\vp_i^{(k)}-\vp_j^{(k)})\big],
\end{align}
and we write
\begin{align}
    & p\Big(\ket{00} \Big| \{c_{ij}^{(\alpha-1)}\}\Big) = p\Big(\ket{11} \Big| \{c_{ij}^{(\alpha-1)}\}\Big) =  \frac{1}{4},\\
    & p\Big(\ket{01} \Big| \{c_{ij}^{(\alpha-1)}\}\Big) = \frac{1}{2}\sum_i |c_{ii}^{(\alpha-1)}|^2 + \frac{1}{4} \sum_{i\neq j} |c_{ij}^{(\alpha-1)}|^2, \\
    & p\Big(\ket{10} \Big| \{c_{ij}^{(\alpha-1)}\}\Big) = \frac{1}{4} \sum_{i\neq j} |c_{ij}^{(\alpha-1)}|^2.
\end{align}
From the protocols analysed previously, we can infer the behavior of this variant. When measuring the bistrings ``00" and ``11", the quantum state remains unchanged. For ``01", we get the same behavior as in the protocol variant with only two auxiliary qubits, i.e.~the coherences $c_{ij}$ where $i\neq j$ are suppressed for every observation of the ``01" bit string. Therefore, this protocol projects the joint quantum register into a Schmidt decomposition form: $\ket{\Psi} = \sum_{i} c_{AB,ii} \ket{\phi_i} \otimes \ket{\phi_i}$.
For the bit string ``10", however, the protocol needs to be reinitialized as this measurement outcome projects the system into the antisymmetric space. The algorithmic overhead due to unsuccessful projection into the symmetric space is the same as for the protocol variant 1.

\clearpage
\section{Generalization to multiple devices}

We discuss how the quantum broadcasting protocol can be extended to more devices. 

\subsection{Projection to the symmetric subspace}

\begin{figure}[htbp]
    \centering
        \begin{tikzpicture}
        \node[scale=1] {
        \begin{quantikz}
            \lstick{\ket{+}} &  \qw &  \qw &  \qw  & \ctrl{1} &  \gate{H} &  \meter{} \arrow[r] & \rstick{$\ket{0}$} \\
             \lstick{\ket{+}} &  \ctrl{3}  & \qw  &  \qw &  \gate[wires=3]{P} & \gate{H} &  \meter{} & \\
             \lstick{\ket{+}} &  \qw  &  \ctrl{3} &  \qw &   & \gate{H} &  \meter{} & \\
             \lstick{\ket{+}} &  \qw  & \qw  &  \ctrl{3} &   & \gate{H}  & \meter{}& \\
            \lstick[wires=3]{$\ket{\Psi^{(k-1)}}$}  & \gate{U(\tau_k)} &  \qw &   \qw &  \qw &  \qw &  \qw\rstick[wires=3]{$\ket{\Psi^{(k)}}$} & \\
              &  \qw &   \gate{U(\tau_k)} &  \qw &  \qw &  \qw &  \qw  & \\
              &  \qw &  \qw &   \gate{U(\tau_k)} &   \qw &  \qw &  \qw  &
        \end{quantikz}};
    \end{tikzpicture}%
    \caption{Circuit for quantum broadcasting with multiple near-perfect target states. $P$ is a projection onto the fully symmetric subspace of the auxiliary qubits and is related to a quantum Schur transformation.}
    \label{fig:ABCcircuit}
\end{figure}
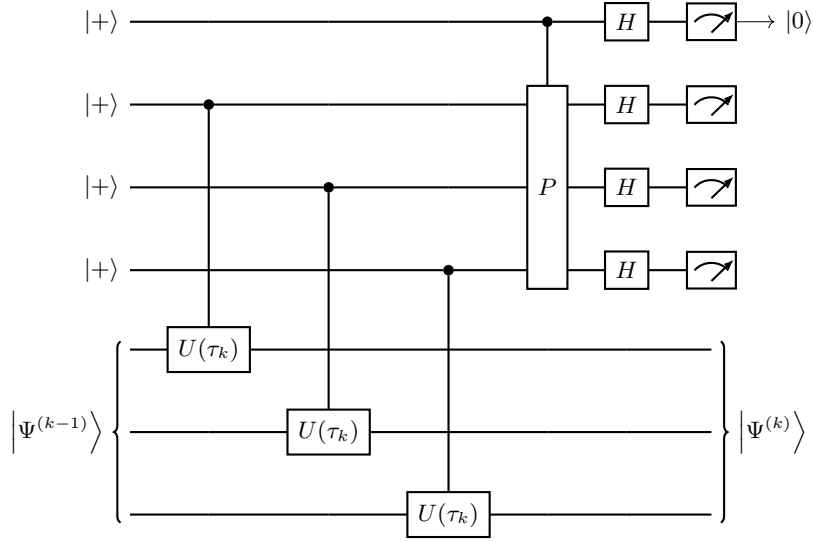

A circuit for three quantum devices that are sharing a limited amount of entanglement through the interaction with one auxiliary qubit each is shown in Fig.~\ref{fig:ABCcircuit}. It is a natural generalization of the two device protocol variant that features a projection to the symmetric subspace through a controlled SWAP operation (and $V=H\otimes H$). Such generalization is done by performing a projection onto the fully symmetric subspace of the auxiliary qubits. Such a projection can be implemented by a (controlled) quantum Schur transform which can be realized with a polynomial number of gates as a function of the number of quantum devices in the protocol~\cite{Bacon2006Efficient}.

We focus the analysis on the bit string measurements ``00\dots0" and ``11\dots1" as they allow for a clearer analytic treatment. For $p$ quantum devices, we analytically compute suppression factors for other bit strings in the near-convergent regime for the quantum broadcasting regime and the results suggest that the ``00\dots0" and ``11\dots1" bit strings indeed lead to the largest suppression, which we conjecture to be optimal. This motivates us to focus on the ``00\dots0" and ``11\dots1" bit strings. 

The process map for these two measurement results is then given as
\begin{align}
    |c_{ij\cdots z}^{(k-1)}|^2\mapsto & |c_{ij\cdots z}^{(k)}|^2 \times \frac{S_{(i,j,\dots,z),\ket{b}}^{(k)}}{N^{(k)}_{\ket{b}}} \\\\
    =& |c_{ij\cdots z}^{(k-1)}|^2\times \left\{
    \begin{array}{ccc}
         \cos(\vp_i^{(k)}/2)^2\cdots \cos(\vp_z^{(k)}/2)^2 /\big(N_{\ket{0\dots0}}^{(k)} \big) &  \mbox{with} & \Pr{}_{\ket{0\dots0}}^{(k)},\\[10pt]
          \sin(\vp_i^{(k)}/2)^2\cdots \sin(\vp_z^{(k)}/2)^2/\big(N_{\ket{1\dots1}}^{(k)} \big) &  \mbox{with} & \Pr{}_{\ket{1\dots1}}^{(k)}.
    \end{array}
    \right.    
\end{align}
The probabilities are
\begin{align} 
    \Pr{}_{\ket{0\dots 0}}^{(\alpha)} &=  \sum_{i,j,\dots,z}  |c_{ij\cdots z}^{(\alpha-1)}|^2 \cos(\vp_i^{(\alpha)}/2)^2 \cos(\vp_j^{(\alpha)}/2)^2\cdots \cos(\vp_z^{(\alpha)}/2)^2 \\ 
    \Pr{}_{\ket{1\dots 1}}^{(\alpha)} &=  \sum_{i,j,\dots,z}  |c_{ij\cdots z}^{(\alpha-1)}|^2 \sin(\vp_i^{(\alpha)}/2)^2 \sin(\vp_j^{(\alpha)}/2)^2\cdots \sin(\vp_z^{(\alpha)}/2)^2.
\end{align}
Now, we shall recall the assumption for the near-convergent regime: we consider the case where we have $|c_d|^2=1$ on all other quantum devices except Bob (i.e.~population only in the dominant eigenstate). On Bob, where we seek to prepare the dominant state from an initial approximation through quantum broadcasting, we have population in the dominant state (amplitude $|c_d|^2$), the subdominant state (amplitude $|c_s|^2$), but the populations of all other eigenstates are zero: $|c_i|^2=0$ for $i\neq d$ and $i\neq s$. Further, we approximate the probability to observe a ``00\dots0" bit string (which is the same as the probability to observe a ``11\dots1" bit string) conditioned on a set of $\{c_{ij\cdots z}^{(\alpha-1)}\}$ as
\begin{align}
      p\Big(\ket{0\dots0} \Big| \{c_{ij\cdots z}^{(\alpha-1)}\}\Big) 
     =& \frac{1}{(2\pi)^N} \int_0^{2\pi} \cdots \int_0^{2\pi} \sum_{i,j,\dots,z}  |c_{ij\cdots z}^{(\alpha-1)}|^2 \cos(\vp_i^{(\alpha)}/2)^2 d\vp_1^{(\alpha)}\dots d\vp_{N}^{(\alpha)} \\
    \rightarrow& \frac{1}{2\pi}\int_0^{2\pi} |c_{dd\cdots d}^{(\alpha-1)}|^2\underbrace{\cos(\vp_d^{(k)}/2)^2 \cos(\vp_d^{(k)}/2)^2 \cdots \cos(\vp_d^{(k)}/2)^2}_{p\ \text{terms}} d\vp_d^{(\alpha)} \\
    =& \frac{\Gamma(p+1/2)}{\sqrt{\pi}\;\Gamma(p+1)} =: F,
\end{align}
with $\Gamma(z)$ the Gamma function ($\Gamma(z)=\int_0^\infty t^{z-1}e^{-t}dt$). Due to symmetries of the trigonometric functions, both bit strings result in the same probability density function. The total probability to observe either all ``00\dots0" or ``11\dots1" in the limit of many devices $p$ is given by a power law
\begin{align}
    \lim_{p \rightarrow \infty} \frac{2\;\Gamma(p+1/2)}{\sqrt{\pi}\;\Gamma(p+1)} = \frac{2}{\sqrt{\pi}} \frac{1}{\sqrt{p}}.
\end{align}
Under the assumptions of a near-convergent regime, we calculate the first moment of $\log \big[ S_{(i,j,\dots,z),\ket{00\dots0}}^{(\alpha)}\big]$ as
\begin{align}
  \mu_{(d,d\dots,d),\alpha,\ket{00\dots0}} =& \mathbb{E}\left[\log \big[ S_{(d,d\dots,d),\ket{00\dots0}}^{(\alpha)} \big] \Big| \ket{00\dots 0}, \{c_{ij\cdots z}^{(\alpha-1)}\} \right]  \\
  &= \frac{1}{2\pi F}\int_0^{2\pi} \log\Big[\underbrace{\cos(\vp_d^{(k)}/2)^2 \cos(\vp_d^{(k)}/2)^2 \cdots \cos(\vp_d^{(k)}/2)^2}_{p\ \text{terms}}\Big] d\vp_d^{(\alpha)} \\
    &= \frac{1}{2\pi F}\int_0^{2\pi} p\log\Big[\cos(\vp_d^{(k)}/2)^{2}\Big] \cos(\vp_d^{(k)}/2)^{2p} d\vp_d^{(\alpha)} \\
    &= p\; [H(p-1/2)-H(p)],
\end{align}
with the harmonic number $H(z):=\int_0^1\frac{1-t^z}{1-t}dt$. Similarly, we compute
\begin{align}
      \mu_{(s,d\dots,d),\alpha,\ket{00\dots0}} =& \mathbb{E}\left[\log \big[ S_{(s,d,\dots,d),\ket{00\dots0}}^{(\alpha)} \big] \Big| \ket{00\dots 0}, \{c_{ij\cdots z}^{(\alpha-1)}\} \right]  \\
  &= \frac{1}{2\pi F}\int_0^{2\pi} \log\Big[\underbrace{\cos(\vp_s^{(k)}/2)^2 \cos(\vp_d^{(k)}/2)^2 \cdots \cos(\vp_d^{(k)}/2)^2}_{p\ \text{terms}}\Big] d\vp_d^{(\alpha)}d\vp_s^{(\alpha)} \\
  &= \frac{1}{2\pi F}\int_0^{2\pi} (p-1)\log\Big[\cos(\vp_d^{(k)}/2)^{2}\Big] \cos(\vp_d^{(k)}/2)^{2p} d\vp_d^{(\alpha)} \\
  &+ \frac{1}{2\pi F}\int_0^{2\pi} \log\Big[\cos(\vp_s^{(k)}/2)^{2}\Big] \cos(\vp_s^{(k)}/2)^{2p} d\vp_d^{(\alpha)} \\
  &= (p-1)[H(p-1/2)-H(p) - \log(4)].
\end{align}
Putting it all together, the average suppression in the limit of large $p$ is
\begin{align}
    \lim_{p\rightarrow\infty} \exp[ \mu_{(s,d\dots,d),\alpha,\ket{00\dots0}} - \mu_{(d,d\dots,d),\alpha,\ket{00\dots0}} ] &=  \lim_{p\rightarrow\infty} \exp[-H(p-1/2)+H(p)-\log(4)] \\
    &=\exp[-\log(4)] = \frac{1}{4},
\end{align}
giving the same average relative suppression for the ``00\dots0" or ``11\dots1" bit strings as the two-copy protocol (Sec.~\ref{sec:twoancillas}) for a ``01" bit string. Note that for $p=2$ we recover the result for the two-copy protocol of $\exp(-7/6)$. 

This result is valid only in the limit of the near-convergent regime. Indeed, numerical analyses of the protocols for two devices suggest that the behavior  protocol where we explicitly project into the symmetric subspace and then measure the auxiliary qubits in the $\sigma^x$-basis (Sec.~\ref{sec:threeancillas}) could behave more favorably than the first protocol (Sec.~\ref{sec:twoancillas}) even though the analytical suppression factor in the near-convergent regime is more favorable for the latter. Therefore, the subspace proposal presented here might find practical applications, even though the analytical suppression value is not suggesting an asymptotic improvement.

\subsection{Comment on a variant using the controlled derangement operator}
We briefly comment on an alternative way to extend the two device protocol to multiple devices. Instead of a (controlled) projection to the fully symmetric subspace, another educated guess for generalizing the protocol would be a controlled derangement operation on the auxiliary qubits~\cite{Koczor2021Exponential}. However, our analysis suggests that this variant does not achieve the desired suppression.

\begin{figure}[htbp]
    \centering
        \begin{tikzpicture}
        \node[scale=1] {
        \begin{quantikz}
            \lstick{\ket{+}} &  \qw &  \qw &  \qw  & \ctrl{1} &  \gate{H} &  \meter{} \arrow[r] & \rstick{$\ket{0}$} \\
             \lstick{\ket{+}} &  \ctrl{3}  & \qw  &  \qw &  \gate[wires=3]{D} & \gate{H} &  \meter{} & \\
             \lstick{\ket{+}} &  \qw  &  \ctrl{3} &  \qw &   & \gate{H} &  \meter{} & \\
             \lstick{\ket{+}} &  \qw  & \qw  &  \ctrl{3} &   & \gate{H}  & \meter{}& \\
            \lstick[wires=3]{$\ket{\Psi^{(k-1)}}$}  & \gate{U(\tau_k)} &  \qw &   \qw &  \qw &  \qw &  \qw\rstick[wires=3]{$\ket{\Psi^{(k)}}$} & \\
              &  \qw &   \gate{U(\tau_k)} &  \qw &  \qw &  \qw &  \qw  & \\
              &  \qw &  \qw &   \gate{U(\tau_k)} &   \qw &  \qw &  \qw  &
        \end{quantikz}};
    \end{tikzpicture}%
    \caption{Circuit proposal for preparing the dominant eigenstate with multiple copies using a derangement operator. Numerical analyses did not suggest that the projection to the fully symmetric space gives better results for quantum broadcasting than the controlled derangement operation.}
    \label{fig:Dcircuit}
\end{figure}
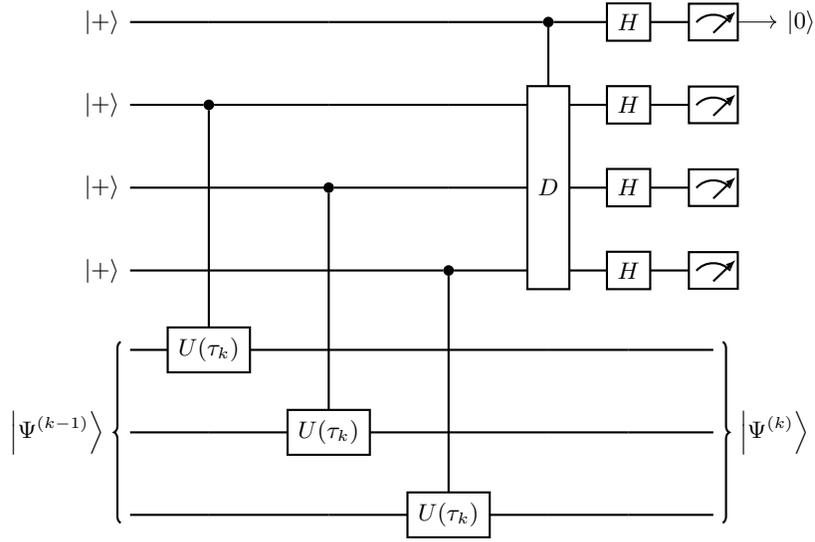

\section{Considerations under the presence of symmetries}\label{ssec:symmetry}
In the main text, we probe the broadcasting protocol numerically for the ZZXZ model $H=\sum_{i=1} (J\sigma_i^z \sigma_{i+1}^z + \sigma_i^x + \sigma_i^z)$ and observed convergence to the ground state. Here, we briefly discuss that under the presence of symmetries in the spectrum of the Hamiltonian, convergence may be hindered and plateau instead of reaching arbitrary precision. However, this issue can be easily addressed. Let us consider the Hamiltonian $H=\sum_{i=1} \sigma_i^z \sigma_{i+1}^z$. The corresponding ground state energy is denoted $\lambda_0$ and the highest excited state energy satisfies $\lambda_{N-1}=-\lambda_0$. We see that the two random variables $\vp_0=\tau\lambda_0$ and $\vp_{N-1}=\tau\lambda_{N-1}$ are perfectly correlated instead of 

\end{document}